\documentstyle[aps,epsf,floats]{revtex}
\newcommand{\beq}{\begin{equation}}
\newcommand{\eeq}{\end{equation}}
\newcommand{\bea}{\begin{eqnarray}}
\newcommand{\eea}{\end{eqnarray}}
\def\({\left(}
\def\){\right)}
\def\Ai{{\rm Ai}}
\def\srx{\sqrt{x}}
\def\non{\nonumber\\}

\newcommand{\lla}{\left\langle}
\newcommand{\rra}{\right\rangle}
 
\begin{document}

\title{Exact statistical properties of the Burgers equation}
\author{L. Frachebourg\footnote{supported by
the Swiss National Foundation for Scientific Research.}
and Ph. A. Martin}
\address{Institut de Physique Th\'eorique}
\address{Ecole Polytechnique F\'ed\'erale de Lausanne}
\address{CH-1015 Lausanne, Switzerland}

\maketitle

\begin{abstract}
 
The one dimensional Burgers equation in the inviscid limit with white
noise initial condition is revisited. 
The one- and two-point distributions of the Burgers field as well as 
the related distributions of shocks are obtained in closed
analytical forms. In particular, the large distance behavior of
spatial correlations of the field is determined.
Since higher order distributions factorize in terms of the one and two
points functions, our analysis provides an explicit and complete
statistical description of this problem.

\end{abstract}

\vskip 1truecm

\section{Introduction} 

The Burgers equation for the velocity field $u(x,t)$
\begin{equation}
{\partial \over \partial t}u(x,t)+u(x,t){\partial \over \partial x}u(x,t)=
\nu {\partial^2\over \partial x^2}u(x,t)
\label{1}
\end{equation}
has recently raised much interest because of its multiple connections to a 
variety of physical and mathematical
problems. 
Background and references can be found for instance in the 
recent book \cite{Woyczynski}.
The original Burgers problem \cite{Burgers}
concerned the statistics of the velocity field $u(x,t)$ and of shock-waves
in the inviscid limit $\nu\to 0$ when the distribution of the initial velocity 
field $u(x,0)$ is a $\delta$-correlated Gaussian (white noise). 
It provides an over-simplified, but analytically tractable model of turbulence
which has attracted a lot of studies over the last decades.
In \cite{Burgers} a considerable amount of work is done 
to calculate various moments 
of these distributions, but the distributions themselves were 
not obtained in closed
forms due to the complexity of the analysis. 
The question has been addressed
again by Tatsumi and Kida \cite{Tat-Kida} and Kida \cite{Kida}.
In \cite{Tat-Kida}, kinetic equations for the dynamics 
of shocks are used to derive 
scaling properties of the distributions, 
and in the second part of \cite{Kida}, 
Kida presents the result of
numerical simulations for the distribution of the strength and 
the velocity of shocks.  
Recently, Avellaneda and E \cite{Ave-E} and Avellaneda \cite{Avellaneda}
have derived rigorous upper and lower bounds 
of the cubic type $\exp(-Cu^{3})$
for the tails of the one-point distribution. 
Such cubic bounds have also been obtained in \cite{Mar-Piasecki}
for the distribution of mass in the closely related 
problem of ballistic aggregation.

In this work we revisit Burgers problem by providing closed analytical 
forms of the statistical distributions for the field and the associated  
distributions of shock-waves. Our main contributions are a simple 
formulae expressing the one-point distribution as integrals over 
the analytic continuation of the Airy
function on the imaginary axis (formulae (\ref{p1u}) and (\ref{pm})) as well
as a detailed study of the clustering behavior of the two-point distributions.
Since it is known that $u(x,t)$ as a function of $x$ (for fixed $t$) is 
a Markov process \cite{Ave-E}, the higher order distributions
factorize in products of one and two-point functions. 
Hence our results give a complete solution 
to the one dimensional Burgers problem
with initial white noise distributed data in the inviscid limit.

In section II, we recall well known facts about the Burgers equation 
in the inviscid limit
with the purpose to introduce the notations and the definitions 
of the one- and two-point
distribution functions. 
In section III, using the notion of first hitting time,
these distributions are expressed in terms of the basic propagator 
for a Brownian motion 
constrained by parabolic barriers. It appears that all statistical properties
of the Burgers problem are embodied in the knowledge of three functions called
here $I$, $J$ and $H$. The functions $I$ and $J$ are calculated in section IV
and the one-point distribution of fields and shocks are discussed. 
In particular
an explicit formula for the strength distribution is obtained. These results
have already been announced in \cite{Frachebourg} in the equivalent language 
of ballistic aggregation. The section V is entirely devoted to the study of 
the large distance behavior of the two-point distribution (the function $H$). 
Since the analysis is somewhat heavy, technical parts have been relegated in 
appendices. Finally the factorization of higher order distributions  
as well as their time dependence are discussed in the conclusion.

The situation considered by Burgers is particularly relevant to the non 
equilibrium statistical model of ballistic aggregation: 
it is known \cite{Burgers,Kida} that the dynamics of shocks in
Burgers turbulence is closely related
to the dynamics of the aggregating particles. White noise initial 
distribution of the
Burgers velocity field corresponds to uncorrelated Maxwellian initial 
velocity distribution 
of the particles undergoing aggregation.
Hence our results also solve this statistical mechanical model. A precise 
connection between the two
problems can only be made in a proper scaling limit since ballistic 
aggregation always retains the
discrete nature of particles whereas Burgers velocity field describes 
a continuous medium. This connection will
be discussed in an other paper \cite{Fra-Mar-Pia}.  
Notice also that the Burgers equation is equivalent to the KPZ equation 
\cite{KPZ} of surface growth processes with the height $h$ of the surface 
given by $u(x,t)=\partial_x h(x,t)$.  
Burgers turbulence arising from other classes of stochastic 
initial data (see {\sl e.g.} \cite{Sinai})
or from the action of external random forces (see {\sl e.g.} \cite{Poliakov}), 
which is the subject
of considerable current investigations, is not discussed in this paper.

\section{General setting}
\label{section2}

For convenience, we shortly recall the 
construction 
of solutions of the Burgers equation in the inviscid limit, see 
\cite{Burgers}, \cite{Woyczynski} and references therein.
Introducing the potential $\partial \Psi(x,t)/\partial x=u(x,t)$
together with the Hopf-Cole transformation
\beq
\Psi(x,t)=-2\nu\ln \theta(x,t),
\eeq
one finds that the function $\theta(x,t)$ satisfies the linear
diffusion equation
\beq
{\partial \over \partial t}\theta(x,t)=
\nu {\partial^2\over \partial x^2}\theta(x,t).
\eeq
It can be readily solved leading to the explicit solution
\beq
u(x,t)={\int_{-\infty}^{\infty}dy {x-y\over t}\exp\left(-{1\over 2\nu}
F(x,y,t)
\right)\over 
\int_{-\infty}^{\infty}dy \exp\left(-{1\over 2\nu}
F(x,y,t)
\right)}
\label{2}
\eeq
where 
\beq
F(x,y,t)={(x-y)^2\over 2t}-\psi(y)
\eeq
with
\beq
\psi(y)=-\Psi(y,0)=-\int_0^y dy^{\prime}\,u(y^\prime,0)
\eeq
which depends upon the initial condition.
Burgers turbulence corresponds here to the situation where the initial 
velocity field $u(x,0)$ is a white noise process in space, or equivalently 
$\psi(y)$ is a two-sided Brownian motion pinned at $\psi(0)=0$.

In the inviscid limit $\nu\to 0$, the only contributions of the integrals in
Eq.(\ref{2})
come from the minima of the function $F(x,y,t)$, which depend on the initial 
condition through $\psi(y)$,
\beq
\xi(x,t)=\min_y F(x,y,t)
\label{3}
\eeq
and we obtain
\beq
u(x,t)={x-\xi(x,t)\over t}.
\label{4}
\eeq

Due to the scaling properties of the solution $u(x,t)$, one can trivially 
take into account the time dependence of the problem.
Indeed the scaled Brownian motion $t^{\alpha/2}\psi(yt^{-\alpha})$ 
is equivalent in probability to $\psi(y)$ so that by (\ref{3}) 
and (\ref{4}) with $\alpha=2/3$ one has that 
$t^{2/3}\xi\left({x\over t^{2/3}},1\right)$ 
is equivalent to $\xi(x,t)$ and $t^{-1/3}u\left({x\over t^{2/3}},1\right)$ 
is equivalent to $u(x,t)$.
We study from now on the fixed time $t=1$ solution $u(x,1)\equiv u(x)$. 
It will then always be possible to recover the time-dependent solution through 
this scaling property as we shall see in the concluding section.

\begin{figure}
\epsfxsize=13truecm
\hspace{3.25truecm}
\epsfbox{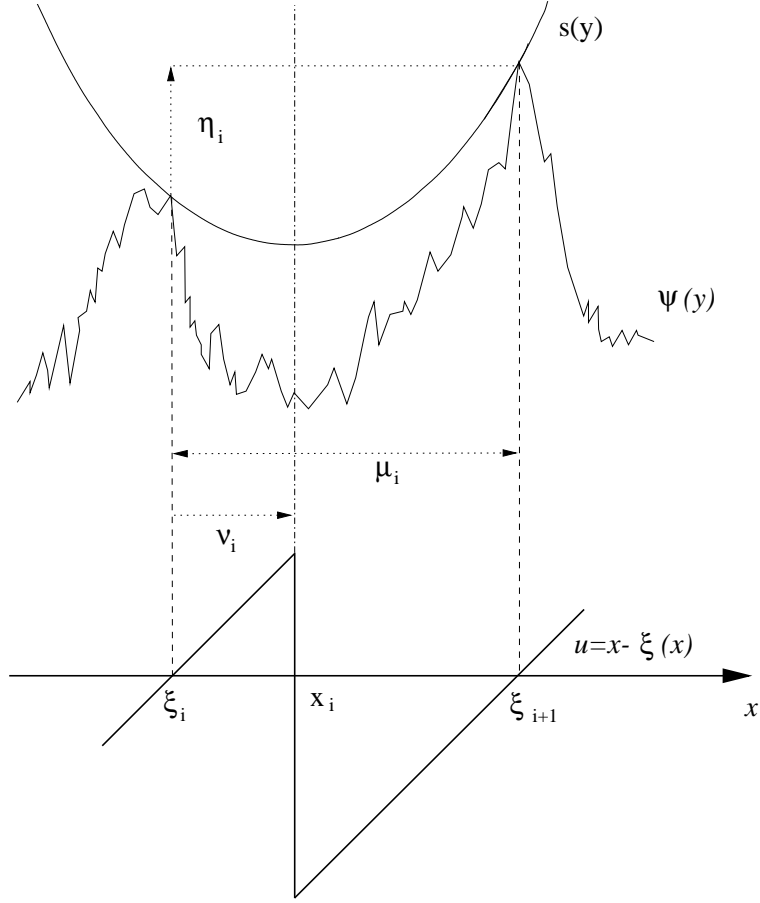}
\caption{
Geometrical interpretation of the solution $u(x)=x-\xi(x)$ for a given 
realization of the Brownian motion $\psi(y)$ which stays below 
a parabola of equation $s(y)=(y-x_i)^2/2+C$ but on  two contact points
$\psi(\xi_i)=(\xi_i-x_i)^2/2+C$ and 
$\psi(\xi_{i+1})=(\xi_{i+1}-x_i)^2/2+C$.
A shock is located at $x_i$ with strength $\mu_i=\xi_{i+1}-\xi_i$ 
and wavelength $\nu_i=x_i-\xi_i$ while 
$\eta_{i}=\mu_{i}^{2}/2-\mu_{i}\nu_{i}$.}
\label{fig1}
\end{figure}

The minimum $\xi(x)\equiv \xi(x,1)$ as 
a function of $x$ can be found with the help
of a nice 
geometrical interpretation of the solution. 
One considers a realization of the Brownian motion $\psi(y)$ 
and a parabola centered at $x$ 
of equation $(x-y)^2/2+C$ (see Fig.\ref{fig1}) and adjusts 
the constant $C$ in order for the parabola to touch $\psi(y)$ 
without ever crossing it. 
The coordinate of the contact point is the minimum $\xi(x)$ leading thus to
$u(x)=x-\xi(x)$. 
Then one glides the parabola on the graph of $\psi(y)$ by 
a continuous change of its center $x$ and $C$ until it 
touches it for $x=x_i$ on two contact points $\xi_{i}$ and $\xi_{i+1}$.
Thus at $x=x_i$, the function $F(x,y,1)$ has two minima  
leading to a discontinuity of $u(x)$, called a shock, where 
$\lim_{\epsilon\to 0}u(x_{i}-\epsilon)
=x_{i}-\xi_{i}$ and $\lim_{\epsilon\to 0}u(x_{i}+\epsilon)=x_{i}-\xi_{i+1}$. 
To make $u(x)$ singled valued at a shock, we define it to be continuous 
from the left setting $u(x_{i})=x_{i}-\xi_{i}$.

A shock is characterized (see Fig.\ref{fig1}) 
by its location $x_{i}$ and two parameters 
which can be taken as\footnote{At time $t$, the strength is usually defined 
as the discontinuity $\mu_i/t=(\xi_{i+1}-\xi_i)/t$ of $u(x,t)$ at a shock.}
\beq
\mu_{i}=\xi_{i+1}-\xi_{i}\;\;\;\;{\rm "strength"},\;\;\;\;
\nu_{i}=x_{i}-\xi_{i}\;\;\;\;\;{\rm "wavelength"}.\label{6}
\eeq
Instead of $\nu_{i}$ it will also be 
convenient to use the parameter $\eta_{i}$
\beq
\eta_{i}
=\frac{\mu_{i}^{2}}{2}-\mu_{i}\nu_{i},\;\;\;\;
\nu_{i}=\frac{\mu_{i}}{2}-\frac{\eta_{i}}{\mu_{i}}.
\label{7}
\eeq

The quantities of interest to be computed are on one hand the joint 
distribution densities $p_n(x_{1},u_{1};x_{2},u_{2};\ldots;x_{n},u_{n})$ 
for the Burgers velocity field to have values in-between 
$u_{1}$ and $u_{1}+du_{1}$, $\ldots$ , 
$u_{n}$ and $u_{n}+du_{n}$ at points $x_{1},\ldots,x_{n}$,
when average is taken over the realizations 
of the initial condition $u(x,0)$.
On the other hand we will also consider 
the joint distribution densities of shocks 
$\rho_n(x_{1},\mu_{1},\eta_{1};x_{2},
\mu_{2},\eta_{2};\ldots;x_{n},\mu_{n},\eta_{n})$.
We shall obtain the joint distribution for the Burgers velocity field $u(x)$
from that of the variable $x-\xi(x)$. At time $t=1$, these two sets of 
variables coincide and we identify both distributions.

Consider first the one-point distribution density $p_1(x,u)$ where 
$u(x)=x-\xi(x)=u$.
Because of translation invariance, $p_1(x,u)=p_1(0,u)\equiv p_1(u)$ and 
$u(0)=-\xi(0)=u$. Hence $p_1(u)$ is the measure of the set 
of all Brownian paths $\psi(y)$ with $\psi(0)=0$ that have their 
first contact (f.c.)\footnote{Consideration of the first contact (or hitting) 
point is consistent with the left continuity of $u(x)$.
If there is a shock at $x_{i}$, 
$u(x_{i})-\lim_{\epsilon\to 0}u(x_{i}+\epsilon)=\xi_{i+1}-\xi_{i}>0$,
implying that $\xi_{i}$ has to be the first contact with the parabola.} 
with a parabola $y^{2}/2+C$ at $\xi(0)=-u$.
As one can set the origin of coordinates 
to be at this contact point, it is given by the measure 
\beq
 p_1(u) = E\{\psi(y)\leq s_{u}(y), y\in R;\; 
\;{\rm f.c. \;with\;}s_{u}(y)\; {\rm at} \;(0,0)\}
\label{7a}
\eeq
of the set of paths that stay below the parabola
\beq
s_{u}(y)=\frac{y^{2}}{2}-uy
\label{7b}
\eeq
and have their first contact with it at $\psi(0)=0$. By first contact in 
(\ref{7a}), we mean that the path is strictly below the parabola 
$\psi(y)< s_{u}(y)$ for $y<0$, is assigned to pass at $\psi(0)=0$ and is 
then such as $\psi(y)\leq s_{u}(y)$ for $y\geq 0$.
The expectation $E\{\cdots\}$ refers to Brownian paths running in
the infinite ``time'' interval $-\infty<y<\infty$.

\begin{figure}
\epsfxsize=16truecm
\hspace{1truecm}
\epsfbox{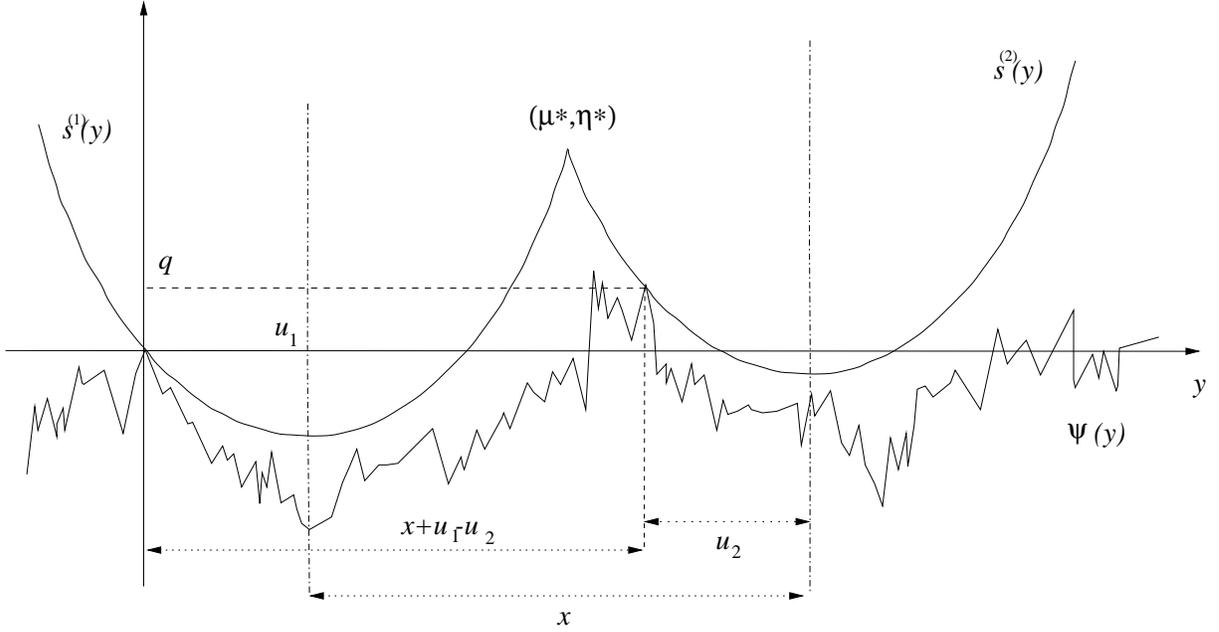}
\caption{
Brownian interpretation of the two-point distribution of 
the velocity field $p_2(x,u_1,u_2)$. 
The Brownian paths stays under the parabolas 
$s^{(1)}(y)=y^2/2-u_1y$ and 
$s^{(2)}(y)=(y-x-u_1)^2/2-u_2^2/2+q$
but on two contact points $\psi(0)=0$ and $\psi(x+u_1-u_2)=q$
where $-(x+u_1-u_2)(x+u_1+u_2)/2\leq q\leq(x+u_1-u_2)(x-u_1-u_2)/2$.
\label{fig2}}
\end{figure}

Likewise, the two-point joint density distribution 
$p_2(0,u_{1};x,u_{2})\equiv p_{2}(x,u_{1},u_{2})$ is the measure of 
the set of paths with $\psi(0)=0$ that have a first contact with a parabola 
$y^2/2+C_1$ (centered at the origin) at $\xi(0)=-u_{1}$  and a first contact
with a second parabola $(y-x)^2/2+C_2$ (centered at $x$) at $\xi(x)=x-u_{2}$. 
Once again, we fix the origin at the contact point with the first parabola. 
Thus $p_2(x,u_{1},u_{2})$ is the measure of the set of 
paths which stay below both the parabolas $s^{(1)}(y)=s_{u_1}(y)$ 
centered at $u_1$  and 
a second parabola $s^{(2)}(y)$ centered at $x+u_1$ of 
equation $(y-x-u_1)^2/2+C$, while the paths
have a first contact point $\psi(0)=0$ with $s^{(1)}(y)$  
and a first contact point $\psi(x+u_1-u_2)=q$ with $s^{(2)}(y)$, 
where $x> 0$, $x+u_1-u_2> 0$, see Fig. \ref{fig2}
\footnote{The case $x+u_1-u_2=0$, i.e. when the two contact points coincide,
is discussed in the next section.}. 
In terms of this parameter $q$ the equation of the second parabola
is
\beq
s^{(2)}(y)=\frac{(y-x-u_1)^2}{2}-\frac{u_{2}^2}{2}+q
\label{7c}
\eeq
Now, $q$ is arbitrary except for the constraints
that the first contact point with $s^{(1)}(y)$ must be below the second 
parabola, namely $s^{(2)}(0)\geq 0$,
and that the first contact point with $s^{(2)}(y)$ must be below the 
first parabola, namely $s^{(1)}(x+u_1-u_2)\geq q$. This leads to the condition 
$-q_1\leq q\leq q_2$ with 
\beq
q_{1}=q_1(x,u_1,u_2)=\frac{1}{2}(x+u_1-u_2)(x+u_1+u_2),
\quad q_{2}=q_2(x,u_1,u_2)=\frac{1}{2}(x+u_1-u_2)(x-u_1-u_2).
\label{7d}
\eeq

Hence
\bea 
&&p_2(x,u_{1},u_{2})=
 E\left\{\psi(y)\leq s^{(1)}(y),\;
\psi(y)\leq s^{(2)}(y),\; y\in R;\right.  \nonumber\\\quad 
&&\quad\quad\left. {\rm f.c.\; with}\; s^{(1)}(y) \;{\rm at}\; (0,0),\; 
{\rm f.c.\; with}\; s^{(2)}(y) \;{\rm at}\; (x+u_{1}-u_{2},q),
\;-q_{1}\leq q \leq q_{2}\right \}.
\label{7e}
\eea
The distributions $p_{1}(u_{1})$ and $p_2(x,u_{1},u_{2})$ 
have the normalizations
\beq
\int_{-\infty}^\infty du_{1}\,p_{1}(u_{1})=1
\label{7f}
\eeq
and
\beq
\int_{-\infty}^\infty du_{2}\,p_2(x,u_{1},u_{2})=p_{1}(u_{1}),\quad
\lim_{x\to 0} p_2(x,u_{1},u_{2})=\delta(u_{1}-u_{2})
\label{7g}
\eeq
The distribution of shocks are defined in the same manner. 
By translation invariance $\rho_1(x,\mu,\eta)=\rho_1(0,\mu,\eta)
\equiv\rho_1(\mu,\eta)$ is independent of $x$. 
It is given by the measure of the set of paths
that have two contacts\footnote{ If a path has more than two contacts
with the parabola, the shock parameters are obtained in terms of the
coordinates of the first and the last contacts.}
with the parabola 
$s_{\nu}(y)=y^{2}/2-\nu y$ 
(recall that $\nu=\frac{\mu}{2}-\frac{\eta}{\mu}$), 
a first contact at $\psi(0)=0$ and 
a last contact (l.c.) at $\psi(\mu)=\eta$,
(see Fig. \ref{fig1})
\beq
\rho_1(\mu,\eta) = E\left\{\psi(y)\leq s_{\nu}(y),
\, y\in R;\;{\rm f.c.\; with}\; s_{\nu}(y)\;{\rm at}\;(0,0);\;
{\rm l.c.\; with}\; s_{\nu}(y)\;{\rm at}\;(\mu,\eta)\right\}.
\label{7h}
\eeq

The joint distribution $\rho_2(0,\mu_{1},\eta_{1};x,
\mu_{2},\eta_{2})$ of two shocks at distance $x$ is the set of paths
that have two contacts with the parabola $s_{\nu_{1}}(y)$ as above and two 
contacts with an other parabola whose characteristics will be given 
in the next section.
Notice that the centers of the two parabolas are separated by a distance $x$.

All quantities will be eventually expressed in terms 
of the transition probability kernel for Brownian motion
in presence of parabolic absorbing barriers \cite{Salminen,Groeneboom}.
Consider the conditional probability density 
$K_\nu(\mu_1,\eta_1,\mu_2,\eta_2)$ for the Brownian motion
$\psi(y)$, starting from $\psi(\mu_1)=\eta_1$, to end at 
$\psi(\mu_2)=\eta_2$ while staying 
under the barrier $\psi(y)<s_\nu(y)=y^2/2-\nu y$ 
for $\mu_1\leq y\leq \mu_2$
\beq
K_\nu(\mu_1,\eta_1,\mu_2,\eta_2)=E_{\mu_1,\eta_1}\left\{
\psi(y)<s_\nu(y),\,\mu_1\leq y\leq \mu_2;\;\psi(\mu_2)=\eta_2 \right\}.
\eeq
It thus satisfies the diffusion equation
\beq
\partial_{\mu_2} K_\nu(\mu_1,\eta_1,\mu_2,\eta_2)
={D\over 2}\partial_{\eta_2}^2
K_\nu(\mu_1,\eta_1,\mu_2,\eta_2)
\label{8}
\eeq
with  $K_{\nu}(\mu,\eta_1,\mu,\eta_2)=\delta(\eta_1-\eta_2)$ and 
$K_{\nu}(\mu_1,s_u(\mu_1),\mu_2,\eta_2)
=K_{\nu}(\mu_1,\eta_1,\mu_2,s_u(\mu_2))=0$.
The parameter $D$ in Eq.(\ref{8}) characterizes the initial condition.
To solve this equation it is convenient to consider the shifted
stochastic process $\phi(y)=\psi(y)-s_\nu(y)$ 
which is a Brownian motion with a parabolic drift. 
Clearly
\beq
K_{\nu}(\mu_{1},\eta_1,\mu_{2},\eta_2)=
\overline{K}(\mu_1,\eta_{1}-s_\nu(\mu_{1}),
\mu_2,\eta_{2}-s_\nu(\mu_{2}))
\label{9}
\eeq
where $\overline{K}$ satisfies the diffusion equation with drift 
\beq
\partial_{\mu_2} \overline{K}(\mu_1,\phi_1,\mu_2,\phi_2)=
s'_\nu(\mu_2)\partial_{\phi_2} \overline{K}(\mu_1,\phi_1,\mu_2,\phi_2)+
{D\over 2}\partial_{\phi_2}^2\overline{K}(\mu_1,\phi_1,\mu_2,\phi_2)
\label{10}
\eeq
with $\overline{K}(\mu,\phi_1,\mu,\phi_2)=\delta(\phi_1-\phi_2)$ and Dirichlet 
boundary conditions $\overline{K}(\mu_1,0,\mu_2,\phi_2)
=\overline{K}(\mu_1,\phi_1,\mu_2,0)=0$.
Eq.(\ref{10}) can be reduced to a diffusion equation with 
linear potential by the transformation
\bea
G(\mu_1,\phi_1,\mu_2,\phi_2)= \overline{K}(\mu_1,\phi_1,\mu_2,\phi_2)
\exp\left[-{1\over D}
\left(\phi_1s'_\nu(\mu_1)-\phi_2s'_\nu(\mu_2)
-{1\over 2}\int_{\mu_1}^{\mu_2}d\mu 
(s'_\nu(\mu))^2\right)\right].
\label{11}
\eea
Then the propagator $G$ is the solution of
the equation 
\beq 
\left({\partial \over \partial \mu_2}
-{D\over 2}{\partial^2\over \partial \phi_2^2}
-{1\over D}\phi_2s_\nu''(\mu_2)\right)G(\mu_1,\phi_1,\mu_2,\phi_2)
=0,\;\;\;\phi_1,\;\phi_2\leq 0
\label{12b}
\eeq
with $G(\mu,\phi_1,\mu,\phi_2)=\delta(\phi_1-\phi_2)$ and Dirichlet 
boundary conditions at the origin,
$G(\mu_1,0,\mu_2,\phi_2)=G(\mu_1,\phi_1,\mu_2,0)=0$. Since $s_\nu''(\mu)=1$, 
this equation can be solved with the help of the spectral 
decomposition of the operator
$-{D\over 2}{\partial^2\over \partial \phi_2^2}-{1\over D}\phi_2$ 
leading to \cite{Burgers,Salminen,Groeneboom}
\beq
G(\mu_1,\phi_1,\mu_2,\phi_2)=
\left({2\over D^2}\right)^{1/3}
\sum_{k\geq 1} {\rm e}^{-\omega_k(\mu_2-\mu_1)/(2D)^{1/3}}
{{\rm Ai}(-(2/D^2)^{1/3} \phi_1-\omega_k)
{\rm Ai}(-(2/D^2)^{1/3} 
\phi_2-\omega_k)\over({\rm Ai}'(-\omega_k))^2}.
\label{13}
\eeq
The Airy function ${\rm Ai}(w)$ \cite{Abra},  solution of
\beq
f''(w)-wf(w)=0,
\label{14}
\eeq
is analytic in the complex $w$ plane, and 
has an infinite countable numbers of zeros $-\omega_k$ on the 
negative real axis, $0<\omega_{1}<\omega_{2}<\cdots$. 

Finally coming back to $K_{\nu}$ with the help of (\ref{9},\ref{11}) 
and introducing the explicit form
(\ref{7b}) of $s_\nu(y)$ leads to
\bea 
&& K_\nu(\mu_1,\eta_1,\mu_2,\eta_2) = 
G(\mu_1,\phi(\mu_{1})),\mu_2,\phi(\mu_{2}))\non
&&\quad\quad\quad\times\exp\left[\frac{1}{D} \left(\phi(\mu_{1})(\mu_{1}-\nu)
-\phi(\mu_{2})(\mu_{2}-\nu)
+\frac{(\mu_{1}-\nu)^{3}}{6}-\frac{(\mu_{2}-\nu)^{3}}{6} \right) \right]
\label{12}
\eea
with $\phi(\mu_{1})=\eta_{1}-s_\nu(\mu_{1})$, 
$\phi(\mu_{2})=\eta_{2}-s_\nu(\mu_{2})$.
Note the symmetry 
$K_\nu(\mu_1,\eta_1,\mu_2,\eta_2)=K_{-\nu}(-\mu_2,\eta_2,-\mu_1,\eta_1)$.

\section{Distributions and transition kernel}
\label{section3}

In this section we relate the distribution functions
to the transition kernel $K_\nu(\mu_1,\eta_1,\mu_2,\eta_2)$. 
We first treat the case of a single first contact point by computing 
the (conditional) probability density
\beq
E_{\mu_{1},\eta_{1}}\{
\psi(y)\leq s_{\nu}(y),\mu_{1}\leq y\leq\mu_{2};
\;{\rm f.c.\; with}\;s_{\nu}(y) \;{\rm at}\;(\mu,s_{\nu}(\mu))
;\;\psi(\mu_2)=\eta_2\}
\label{2.1}
\eeq
that a Brownian
motion starting at point $(\mu_1,\eta_1)$ ends at $(\mu_2,\eta_2)$ while
staying below the parabola $s_\nu(y)$, and has a first contact point 
$\psi(\mu)=s_\nu(\mu)$ at "time" ${\mu}$, with $\mu_1<\mu<\mu_2$ and 
$\eta_1<s_\nu(\mu_1)$ and $\eta_2<s_\nu(\mu_2)$.
This enables us to write the probability $p_1(u)$ (\ref{7a}), 
where the expectation is taken on
paths that run in the whole "time" interval $y\in R$, as 
\bea
p_1(u) &=& 
\lim_{\mu_1\to -\infty}\lim_{\mu_2\to\infty}
\int_{-\infty}^{s_u(\mu_1)}d\eta_1
\int_{-\infty}^{s_u(\mu_2)}d\eta_2\non 
&&\quad\quad E_{\mu_{1},\eta_{1}}
\{\psi(y)\leq s_{u}(y),\mu_{1}\leq y\leq\mu_{2};\;
{\rm f.c.\; with}\;s_{u}(y)\; {\rm at}\;(0,0);\;\psi(\mu_2)=\eta_2\}
\label{2.1a}
\eea
As in the preceding section it is convenient to consider 
\beq
E_{\mu_{1},\phi_1}\{\phi(y)\leq 0,
\mu_{1}\leq y\leq\mu_{2},\;{\rm f.c.\; with\; the\; origin\; at}\;(\mu,0);\;
\phi(\mu_2)=\phi_2\}
=-\partial_{\mu} P_{\mu_1,\phi_1;\mu_2,\phi_2}(\mu)
\label{2.2}
\eeq
the quantity corresponding to (\ref{2.1}) for 
the shifted process $\phi(y)=\psi(y)-s_{\nu}(y)$. It is the (conditional)
probability that a drifted Brownian motion $\phi(\mu)$, starting at 
 point $(\mu_1,\phi_1)$, ends at $(\mu_2,\phi_2)$, 
stays negative $\phi(y)\leq 0$
and has a first contact with the origin at "time" $\mu$ ($\phi(\mu)=0$).
The desired quantity 
(\ref{2.1}) is obtained by setting 
$\phi_1=\phi(\mu_{1})=\eta_{1}-s_{\nu}(\mu_{1}),\phi_2
=\phi({\mu_{2}})=\eta_{2}-s_{\nu}(\mu_{2})$ 
in (\ref{2.2}). We have also written that (\ref{2.2}) 
is the density of the probability 
$P_{\mu_1,\phi_1;\mu_2,\phi_2}(\mu)$ that, under the same constraints, 
the path  has its first contact with the origin 
at some "time" larger or equal to $\mu$. 
This probability is given by (for basic notions on first hitting time see 
\cite{Feller})
\beq
P_{\mu_1,\phi_1;\mu_2,\phi_2}(\mu)=\int_{-\infty}^0 d\phi\,
\overline{K}(\mu_1,\phi_1,\mu,\phi)\left[-\partial_\phi\overline{K}
(\mu,\phi,\mu_2,\phi_2)
-\partial_{\phi_2}\overline{K}(\mu,\phi,\mu_2,\phi_2)\right]
\label{2.3}
\eeq
Indeed one considers the paths starting from $(\mu_1,\phi_1)$ that stay 
negative up to $(\mu,\phi)$ and then vanish at some ``time'' 
larger or equal to $\mu$. The probability density for the later
part is given by the measure of paths staying below 
the displaced barrier $\phi(y)<\epsilon$ diminished by that of
paths staying below the origin
$\phi(y)<0$ as $\epsilon\to 0$,  namely by
\beq
\lim_{\epsilon\to 0}\frac{1}{\epsilon}
(\overline{K}(\mu,\phi+\epsilon,\mu_2,\phi_2+\epsilon)
-\overline{K}(\mu,\phi,\mu_2,\phi_2))
\label{2.4}
\eeq
This leads to (\ref{2.3}).
Introducing (\ref{2.3}) in (\ref{2.2}) and using the forward diffusion equation
(\ref{10}) as well as its backward equivalent, 
we find after several integrations by parts that
\bea
E_{\mu_{1},\phi_1}\{\phi(y)\leq 0,\;
\mu_{1}\leq y\leq\mu_{2};&&\;{\rm f.c. \;with\;the\;origin\;at}\;(\mu,0);\;
\phi(\mu_2)=\phi_2\}\non
&&={D\over 2} \partial_\phi 
\overline{K}(\mu_1,\phi_1,\mu,\phi)\partial_\phi
\overline{K}(\mu,\phi,\mu_2,\phi_2)\bigg|_{\phi=0}
\label{2.5}
\eea
Coming back to the original variables our probability (\ref{2.1}) reads
\bea
E_{\mu_{1},\eta_{1}}\{
\psi(y)\leq s_{\nu}(y),\;\mu_{1}\leq y\leq\mu_{2};&&\;{\rm f.c. \; with}\;
s_{\nu}(y)\; {\rm at}\;(\mu,s_{\nu}(\mu));\;\psi(\mu_2)=\eta_2\}\nonumber\\
&& ={D\over 2}
\partial_\eta K_\nu(\mu_1,\eta_1,\mu,\eta)
\partial_\eta K_\nu(\mu,\eta,\mu_2,\eta_2)\bigg|_{\eta=s_\nu(\mu)}
\label{2.6}
\eea
with $K_\nu$ given by Eq.(\ref{12}).

Let us define the function $J(\nu)$ to be
\beq
J(\nu)=-\sqrt{{D\over 2}}
\lim_{\mu_2\to\infty}\int_{-\infty}^{s_\nu(\mu_2)}d\eta_2 
\partial_\eta K_\nu(0,\eta,\mu_2,\eta_2)\bigg|_{\eta=0}.
\label{2.8}
\eeq
Then, from (\ref{2.6}) and (\ref{2.1a}) it is now straightforward to 
find the expression of the
one-point distribution of the velocity field
\beq
p_1(u)= J(-u)J(u)
\label{2.7}
\eeq
where we used the fact 
that $K_\nu(\mu_1,\eta_1,0,\eta)=K_{-\nu}(0,\eta,-\mu_1,\eta_1)$.

We come now to the two-point function (\ref{7e}) which involves a first 
contact at $y=0$ with the parabola $s^{(1)}(y)$ and a 
first contact at $y=x+u_{1}-u_{2}$
with the second parabola $s^{(2)}(y)$. We consider first the situation
where these two contact points are distincts, i.e. 
when the strict inequality    
$x+u_{1}-u_{2}>0$ holds. Since $u(x)$ has slope
equal to one except at the location of shocks, this corresponds 
to velocity fields with $u(0)=u_{1}, u(x)=u_{2}$ 
that have at least one shock in the interval $[0,x)$.   

When $x+u_{1}-u_{2}>0$, each contact gives rise to an
expression of the form (\ref{2.6}) with appropriate parameters 
(see Fig. \ref{fig2}). The first contact with the parabola
$s^{(1)}(y)=s_{u_{1}}(y)$ is as before. After the "time" $\mu^{*}$
(coordinate of the parabolas intersection $s^{(1)}(\mu^*)=s^{(2)}(\mu^*)$),
the paths are found under the second parabola
$s^{(2)}(y)$ (\ref{7c}) with corresponding
propagator $K_{s^{(2)}}(\mu_{1},\eta_{1},\mu_{2},\eta_{2})$ 
and first contact at $(x+u_{1}-u_{2},q)$. 
The corresponding probability is given by the following arrangement
\bea
&&{D\over 2}
\partial_\eta K_{u_{1}}(\mu_1,\eta_1,0,\eta)
\partial_\eta K_{u_{1}}(0,\eta,\mu^{*},\eta^{\prime})\bigg|_{\eta=0}\nonumber\\
&&\quad \times {D\over 2}
\partial_\eta K_{s^{(2)}}(\mu^{*},\eta^{\prime},x+u_{1}-u_{2},\eta)
\partial_\eta K_{s^{(2)}}(x+u_{1}-u_{2},\eta,\mu_2,\eta_2)
\bigg|_{\eta=q}
\label{2.8a}
\eea
which has to be integrated on $\eta_{1},\eta_{2},\eta^{\prime}$ and 
$q$ in the appropriate ranges
and taken in the limits $\mu_{1}\to-\infty,\;\mu_{2}\to\infty$.
The propagator associated with the second parabola can be written
in a coordinate system where the second contact point is again located 
at the origin, namely 
\beq
K_{s^{(2)}}(\mu_{1},\eta_{1},\mu_{2},\eta_{2})
=K_{u_{2}}(\mu_{1}-x-u_{1}+u_{2},\eta_{1}-q,\mu_{2}-x-u_{1}+u_{2},\eta_{2}-q)
\label{2.9}
\eeq
One finds finally
\beq
p_2(u_{1},u_{2},x)=J(-u_1)H(x,u_1,u_2)J(u_2), \;\;\; x+u_{1}-u_{2}>0
\label{2.10}
\eeq
where the function $H(x,\nu_1,\nu_2)$ is defined as
\beq
 H(x,\nu_1,\nu_2)=
{D\over 2}\int_{-q_1}^{q_2}dq \int_{-\infty}^{\eta^*} 
d\eta^{\prime} \,\partial_\eta K_{\nu_1}(0,\eta,\mu^*,\eta^{\prime})
\bigg|_{\eta=0}
\partial_\eta K_{\nu_2}(\mu^{*}-x-\nu_{1}+\nu_{2},\eta^{\prime}-q,0,\eta)
\bigg|_{\eta=0}.
\label{2.11}
\eeq
The integration limits  $q_{1}=q_1(x,\nu_1,\nu_2)$ and 
$q_{2}=q_2(x,\nu_1,\nu_2)$ are given by (\ref{7d}). 
The intersection point between the two parabolas has coordinate 
$(\mu^*,\eta^*)=((q+q_1)/x,\mu^{*2}/2-\nu_1\mu^*)$. 

We now determine the contribution to $p_2(x,u_{1},u_{2})$ of the set 
of velocity fields $u(x)$ that have no shocks in $[0,x)$ 
(i.e. when $x+u_{1}-u_{2}=0$) with the help of the normalization (\ref{7g}). 
The set of Burgers fields with $u(0)=u_{1}$ can be divided
into the union of two disjoint sets, those having at least one shock in  
$[0,x)$ and those having no shocks in  $[0,x)$. As seen before the first 
set corresponds to Brownian paths having two distinct contact points and from 
the previous discussion its measure is given by 
$\int_{-\infty}^{u_{1}+x}du_{2}J(-u_1)H(x,u_1,u_2)J(u_2)$. The second 
set corresponds to the case $x+u_{1}-u_{2}=0$ when Brownian paths have a 
first contact point $\psi(0)=0$ at the intersection of the two parabolas 
$s_{u_{1}}(y)$ and $s_{u_{1}+x,}(y)$
with measure 
\beq
E\{\psi(y)\leq s_{u_{1}}(y),\; y< 0;\;\psi(y)\leq s_{u_{1}+x}(y),\;y\geq 0;
\;{\rm f.c. \; with}\; s_{u_1}(y)\; {\rm at}\; (0,0)\}=J(-u_{1})J(u_{1}+x)
\label{2.11a}
\eeq
The result (\ref{2.11a}) is derived by a slight extension of the calculation 
that led to (\ref{2.7}). The measures of these two sets sum up to $p_1(u_1)$
\beq
J(-u_{1})J(u_{1}+x)+\int_{-\infty}^{u_{1}+x}du_{2}J(-u_1)H(x,u_1,u_2)J(u_2)
=p_{1}(u_{1}).
\label{2.11b}
\eeq
Hence we conclude from (\ref{7g}) that the complete form of 
$p_{2}(x,u_{1},u_{2})$ is 
\beq
p_{2}(x,u_{1},u_{2})=J(-u_{1})\left[\delta(x+u_{1}-u_{2})
+\theta(x+u_{1}-u_{2})H(x,u_1,u_2)\right] J(u_2).
\label{2.11c}
\eeq
A quantity of interest is the probability density $p_{[0,x)}(u_{1})$ for the 
Burgers field to take the value $u_{1}$ at $x=0$ while there is no shock 
in the interval $[0,x)$, i.e. $u(x)=u_{1}+x$.
This is precisely the quantity (\ref{2.11a}), namely integrating (\ref{2.11c})
on $u_2$ with $H$ omitted
\beq
p_{[0,x)}(u_{1})=J(-u_{1})J(u_{1}+x)
\label{2.11d}
\eeq
and thus
\beq
p_{[0,x)}=\int_{-\infty}^\infty du_{1}J(-u_{1})J(u_{1}+x)
\label{2.11e}
\eeq
is the distribution of intervals of length $x$ without shocks.

We turn now to the shocks distribution functions. According to the discussion 
of previous section Eq.~(\ref{7h}), we use Eq.~(\ref{2.6}) 
to write the one-shock distribution function considered as a 
function of the parameters $\mu,\eta$ (see Fig. \ref{fig1})
\beq
\rho_1(\mu,\eta)=J(-\nu)I(\mu,\eta)J(-\mu+\nu)=
J\(-\frac{\mu}{2}+\frac{\eta}{\mu}\)I(\mu,\eta)J\(
-\frac{\mu}{2}-\frac{\eta}{\mu}\)
\label{2.13}
\eeq
where the function $I(\mu,\eta)$ is defined as
\beq
I(\mu,\eta)={D\over 2}\partial_{\eta_1}
\partial_{\eta_2} K_\nu(0,\eta_1,\mu,\eta_2)\bigg|_{\eta_1=0,\eta_2=\eta}.
\label{2.14}
\eeq

The two-shocks distribution $\rho_2(0,\mu_{1},\eta_{1};x,\mu_{2},\eta_{2})
\equiv\rho_2(x;\mu_{1},\eta_{1};\mu_{2},\eta_{2})$ 
(considered as a function of the shock parameters 
$\eta_{1},\mu_{1}$ and $\eta_{2},\mu_{2}$) can be written
as 
\bea
&&\rho_2(x,\mu_{1},\eta_{1},\mu_{2},\eta_{2})=J(-\nu_{1})I(\mu_{1},\eta_{1})
\left[\delta(x+\nu_{1}-\nu_{2}-\mu_{1})\right.\non
&&\quad \quad \left.+\theta(x+\nu_{1}-\nu_{2}-\mu_{1})H(x,-\mu_1+\nu_1,\nu_2)
\right] I(\mu_2,\eta_2)J(-\mu_2+\nu_2)
\label{2.15}
\eea
with $\nu_i=\mu_i/2-\eta_i/\mu_i$, $i=1,2$, and  
the functions $I$, $J$ and $H$ are defined above.

We denote $\rho^{(nn)}_{2}(x;\mu_{1},\eta_{1};\mu_{2},\eta_{2})$ 
the probability density of two nearest neighbors shocks separated by a 
distance $x$; $\rho^{(nn)}_{2}(x;\mu_{1},\eta_{1};\mu_{2},\eta_{2})$ 
is given by the formula (\ref{2.15})
with the $H$ function omitted. 
Then the conditional probability density 
$\rho^{(nn)}(\mu_{1},\eta_{1}|x,\mu_{2},\eta_{2})$ that given a shock 
$\mu_{1},\eta_{1}$ at $x=0$, the next shock $\mu_{2},\eta_{2}$ occurs at $x>0$
is found to be
\bea
\rho^{(nn)}(\mu_{1},\eta_{1}|x,\mu_{2},\eta_{2})&=& 
\frac{\rho^{(nn)}_{2}(x;\mu_{1},\eta_{1};\mu_{2},\eta_{2})}
{\rho_1(\mu_{1},\eta_{1})}\non
&=&\delta\(x-\frac{\eta_{1}}{\mu_{1}}+\frac{\eta_{2}}{\mu_{2}}
-\frac{\mu_{1}+\mu_{2}}{2}\)
\frac{I(\mu_{2},\eta_{2})J(-\frac{\mu_{2}}{2}-\frac{\eta_{2}}{\mu_{2}})}
{J(-\frac{\mu_{1}}{2}-\frac{\eta_{1}}{\mu_{1}})}
\label{2.16a}
\eea
This conditional probability has the normalization 
\beq
\int_{0}^{\infty}dx\int_{0}^{\infty}d\mu_{2}
\int_{-\infty}^{\infty}d\eta_{2}\,
\rho^{(nn)}(\mu_{1},\eta_{1}|x,\mu_{2},\eta_{2})=1
\eeq
which leads to the following integral relation between the functions 
$I$ and $J$
\beq
J(\nu)=\int_{0}^{\infty}d\mu\int_{-\infty}^{\infty}d\eta\;
\theta\(\frac{\mu}{2}-\frac{\eta}{\mu}-\nu\) 
I(\mu,\eta)J\(-\frac{\mu}{2}-\frac{\eta}{\mu}\)
\label{2.17a}
\eeq

This analysis shows that the one-point and the two-point distribution 
functions of the Burgers velocity field $u(x)$ 
as well as of the statistics of shocks are entirely determined 
by the knowledge 
of three functions $I$, $J$ and $H$ defined in 
(\ref{2.14}), (\ref{2.8}) and (\ref{2.11}). 
Finally, these last three functions can be computed
from the basic transition kernel
$K_{\nu}$ given by Eq.(\ref{12}).

\section{The functions $I$ and $J$ and the one-point distribution}
\label{section4}

In this section we give explicit expressions for the functions $I(\mu,\eta)$
and $J(\nu)$ defined respectively by Eqs.(\ref{2.14}) and (\ref{2.8}). Through
Eqs.~(\ref{2.7},\ref{2.13}), we will then
obtain explicit forms for the one-point distribution function
of the velocity field $p_1(u)$ and of the shocks $\rho_1(\mu,\eta)$. 

Using the form (\ref{12}) 
of the transition density $K_\nu$ in Eq.(\ref{2.14}) we have that
\beq
I(\mu,\eta)=2a^3
\exp\left(-a^3\left[{\eta^2\over \mu}+{\mu^3\over 12}\right]\right)
{\cal I}(\mu).
\label{iii}
\eeq
We set $a=(2D)^{-1/3}$ and
\beq
{\cal I}(\mu)=\sum_{k\geq 1} {\rm e}^{-a\omega_k \mu}
\label{cali}
\eeq
where $-\omega_k$, $k\geq 1$,  are the zeroes of the Airy function.
This last expression has already been found by Burgers \cite{Burgers}. 
Our point here is to give
a closed form for the function $J(\nu)$ and thus 
for the one-point distributions. 
Inserting (\ref{12}) in (\ref{2.8}) and changing the 
variable $x=-(2/D^2)^{1/3}(\eta_2-s_\nu(\mu_2))$ leads to
\beq
J(\nu)=\sqrt{a}\lim_{\mu\to \infty}{\rm e}^{-a^3[(\mu-\nu)^3+\nu^3]/3}
\int_0^\infty dx\, {\rm e}^{ax(\mu-\nu)}
\sum_{k\geq 1}{\rm e}^{-a\omega_k\mu}{{\rm Ai}(x-\omega_k)
\over {\rm Ai}'(-\omega_k)}.
\label{jj1}
\eeq
It is convenient to introduce the following integral representation
of the sum for $\mu>0$
\beq  
\sum_{k\geq 1}{\rm e}^{-a\omega_k\mu}{{\rm Ai}(x-\omega_k)
\over {\rm Ai}'(-\omega_k)}={1\over 2\pi i}\int_{{\cal C}}dw\,
{\rm e}^{aw \mu}{{\rm Ai}(w+x)
\over {\rm Ai}(w)}
={1\over 2\pi i }\int_{-i\infty}^{i\infty}dw\,
{\rm e}^{aw \mu}{{\rm Ai}(w+x)
\over {\rm Ai}(w)}
\label{jj2}
\eeq
where the contour ${\cal C}$ 
runs just above and below the negative real $w-$axis 
encircling the zeros of the Airy function. 
From the asymptotics  
\beq
\Ai(w)= (4\pi\sqrt {w})^{-1/2}e^{-2w^{3/2}/3}
\left(1+{\cal O}(w^{-3/2})\right),
\;\;\;|w|\to\infty,\;\;\;|{\rm arg}\,w|<\pi
\label{jj3}
\eeq
one deduces that for $w=|w|e^{i\theta}$, $\pi/2\leq \theta<\pi$, one has
$|\Ai(w+x)/\Ai(w)|\sim \exp\(-x|w|^{1/2}\cos(\theta/2)\)$, $|w|\to\infty$,
$\cos(\theta/2)>0$. For $\theta=\pi$, the factor ${\rm e}^{aw \mu}$ ensures 
the convergence in (\ref{jj2}).
Hence for $\mu>0$ one can deform the contour ${\cal C}$ and show that
the unique contribution to the integral comes from the imaginary axis 
$-i\infty<w<i\infty$ leading to the last part of the identity 
(\ref{jj2}).
After exchange of the integrations order one finds 
\beq
J(\nu)=\sqrt{a}\lim_{\mu\rightarrow \infty}{\rm e}^{-a^3[(\mu-\nu)^3+\nu^3]/3}
{1\over 2i\pi }\int_{-i\infty}^{i\infty}dw
{{\rm e}^{a \mu w}
\over {\rm Ai}(w)}
\int_0^\infty dx\, {\rm e}^{-ax(\nu-\mu)}{\rm Ai}(x+w).
\label{jj}
\eeq 
To proceed
we determine first the Laplace transform of $f(x)={\rm Ai}(x+w)$, $w$ fixed,
\beq
\tilde f(s)=\int_0^\infty dx\, {\rm e}^{-xs}f(x).
\eeq
The function $f(x)$ is solution of the second-order differential equation
\beq
f''(x)-(x+w)f(x)=0
\eeq
with $f(0)={\rm Ai}(w)$ and $f'(0)={\rm Ai}'(w)$.
The Laplace transform of this equation is
\beq
\tilde f'(s)+(s^2-\omega)\tilde f(s)=sf(0)+f'(0).
\eeq
with solution 
\beq
\tilde f(s)=\left(\tilde f(0)+\int_0^s d\sigma\,(\sigma f(0)+f'(0))
{\rm e}^{-w \sigma+\sigma^3/3}\right){\rm e}^{w s-s^3/3}
\eeq
\beq
\tilde f(0)=\int_0^\infty dx\, {\rm Ai}(x+w)
=-\pi\left[{\rm Ai}'(w){\rm Gi}(w)
-{\rm Ai}(w){\rm Gi}'(w)\right]
\eeq
where ${\rm Gi}(w)=\pi^{-1}\int_0^\infty dt\,\sin(t^3/3+w t)$ 
\cite{Abra}.

Inserting this Laplace transform in Eq.(\ref{jj}) and using various 
properties of the Airy functions \cite{Abra} leading to the identity 
\beq
\tilde f(0)-\int_{-\infty}^0 d\sigma\,(\sigma f(0)+f'(0))
{\rm e}^{-\omega \sigma+\sigma^3/3}=1,
\eeq
we eventually find
\beq
J(\nu)=\sqrt{a}{\rm e}^{-a^3\nu^3/3}{\cal J}(\nu)
\label{jjj}
\eeq
with
\beq
{\cal J}(\nu)={1\over 2i\pi }\int_{-i\infty}^{i\infty}dw
{{\rm e}^{a\nu w}\over {\rm Ai}(w)}.
\label{calj}
\eeq
Note that this integral is convergent for positive and negative $\nu$.

\begin{figure}
\epsfxsize=11truecm
\hspace{2.75truecm}
\epsfbox{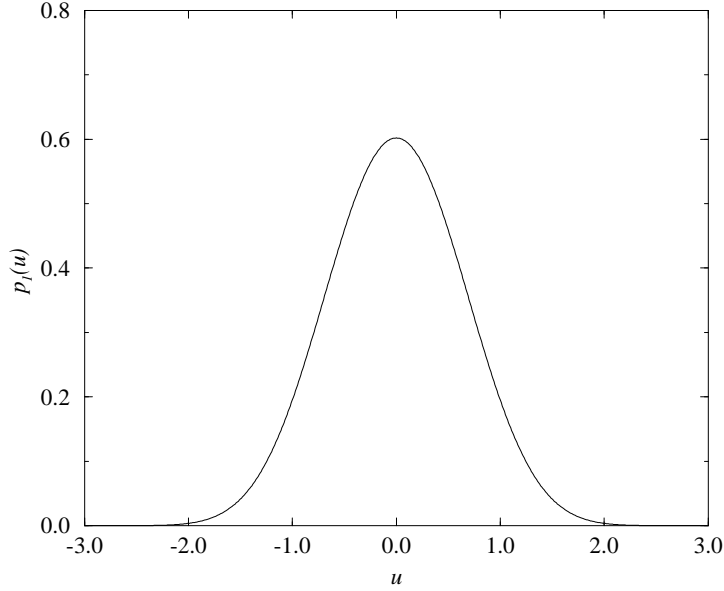}
\caption{The one-point distribution function (\ref{p1u}) 
for the velocity field 
$p_1(u)$ as a function of $u$ for $D=1/2$ ($a=1$).
\label{fig3}}
\end{figure}

With the explicit form (\ref{jjj}) 
of the function $J(\nu)$, the one-point
distribution function  $p_1(u)$ of the velocity field is given by
\beq
p_1(u)=J(u)J(-u)=a{\cal J}(u){\cal J}(-u)
\label{p1u}
\eeq
which is plotted on Fig. \ref{fig3}.

Defining the moments of the distribution as 
$\langle u^n\rangle=\int du\,u^n p_1(u)$ 
we have $\langle u\rangle=0$ as $p_1(u)=p_1(-u)$ and 
$\langle u^2\rangle=m_1(D/2)^{2/3}$ with a constant $m_1\simeq 1.054$.
The normalization (\ref{7f}) is verified as, from (\ref{p1u}),  
\beq
\int_{-\infty}^{\infty} du\, p_1(u)={1\over 2i\pi}\int_{-i\infty}^{i\infty}
{dw \over [{\rm Ai}(w)]^2}
\eeq
can be shown to be equal to one.

To determine the asymptotic behaviour of $p_{1}(u)$, we remark 
that for positive $u$, 
we can close the contour in (\ref{calj}) to encircle 
the poles of the integrand
and thus express 
${\cal J}(u)$ as a sum on the zeros of the Airy function
\beq
{\cal J}(u)=\sum_{k\geq 1} {{\rm e}^{-au\omega_k}\over {\rm Ai}'(-\omega_k)},
\quad u>0
\eeq
Hence ${\cal J}(u)
\sim {\rm e}^{-au\omega_1}/{\rm Ai}'(-\omega_1)$  as $u\to \infty$.
The behavior  of ${\cal J}(u)$ for $u\to -\infty $ can be determined with 
Laplace method to be ${\cal J}(u)\sim -2au\exp(a^3u^3/3)$ and so the large 
$|u|$ behavior of $p_1(u)$ reads
\beq
p_1(u)\sim {2a^2|u|\over {\rm Ai}'(-\omega_1)}
\exp\left(-{a^3 |u|^3\over 3}-a|u|\omega_1\right),\quad |u|\to\infty.
\label{asympp1}
\eeq
This result is of course compatible with the bounds found in Th. 1 of 
\cite{Avellaneda}, but cubic bounds cannot be saturated because of the 
additional exponential decay $\exp(-a|u|\omega_1)$.
It is interesting to remark that, starting form a Gaussian distributed 
initial velocity field $u(x,0)$, 
the field immediately evolves to a distribution which is
not Gaussian but behaves as Eq.(\ref{asympp1}).

Let us turn now to the one-shock distribution function $\rho_1(\mu,\eta)$.
Collecting results Eqs.~(\ref{2.13},\ref{iii},\ref{jjj}), we find
\beq
\rho_1(\mu,\eta)=J\left({\eta\over\mu}-{\mu\over 2}\right)I(\mu,\eta)
J\left(-{\eta\over\mu}-{\mu\over 2}\right)=
2a^4 {\cal J}\left({\eta\over\mu}-{\mu\over 2}\right)
{\cal I}(\mu){\cal J}\left(-{\eta\over\mu}-{\mu\over 2}\right)
\label{distrib}
\eeq
with ${\cal I}$ and ${\cal J}$ defined in (\ref{cali}) and (\ref{calj}), 
respectively.

\begin{figure}
\epsfxsize=11truecm
\hspace{2.75truecm}
\epsfbox{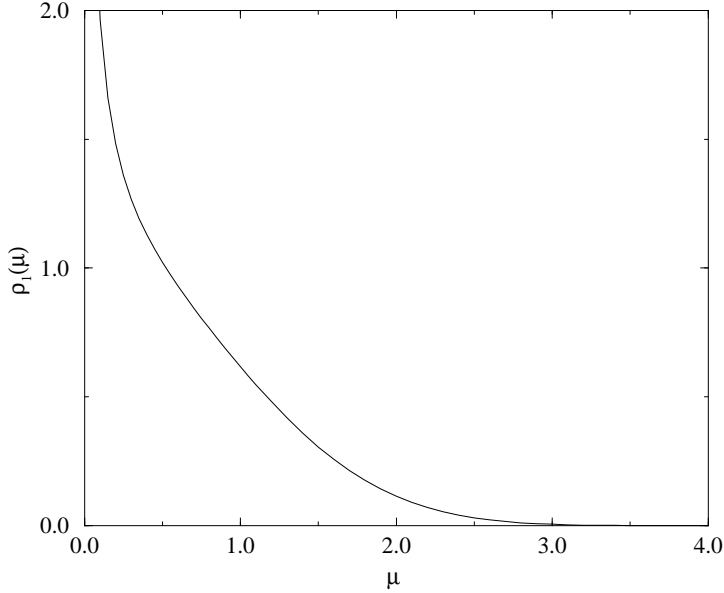}
\caption{
Shock strength distribution $\rho_1(\mu)$ for $D=1/2$, 
($a=1$ in Eq.(\ref{pm})).
\label{fig4}}
\end{figure}

One can compute the shock strength distribution defined as
\beq
\rho_1(\mu)=\int_{-\infty}^\infty d\eta\,\rho_1(\mu,\eta).
\eeq
Inserting (\ref{distrib}) in this last equation,  
we find after the change of variables $w=i\zeta$ and
$\eta^{\prime}=a\eta/\mu$
\beq
\rho_{1}(\mu)=2a^{3}\mu {\cal I}(\mu){1\over (2\pi)^2}
\int_{-\infty}^{\infty}d\zeta_{1}
\int_{-\infty}^{\infty}d\zeta_{2}
\frac{e^{-ia\mu(\zeta_{1}+\zeta_{2})/2}}{\Ai(i\zeta_{1})\Ai(i\zeta_{2})}
\int_{-\infty}^{\infty}d\eta^{\prime}e^{i\eta^{\prime}(\zeta_{1}-\zeta_{2})}
\eeq
which reduces to
\beq
\rho_1(\mu)=2a^3\mu{\cal I}(\mu){\cal H}(\mu)
\label{pm}
\eeq
with 
\beq
{\cal H}(\mu)={1\over 2i\pi }\int_{-i\infty}^{i\infty}
dw{{\rm e}^{-a\mu w}
\over {\rm Ai}^2(w)}.
\eeq
The form of the shock strength 
distribution (\ref{pm}) is plotted on Fig. \ref{fig4}.
Notice that $L\lla \mu\rra$ is the space covered by the shock strength in
a box of size $L$; it is equal to $L$ and one has thus
$\int_0^\infty d\mu\,\mu \rho_1(\mu)=1$. 

We can now determine the behavior of the shock strength distribution for small
and large shocks. For $0<\mu\ll 1$, we use the normalization condition 
(\ref{7f}) to find ${\cal H}(\mu)=1+{\cal O}(\mu)$ while the behavior 
of ${\cal I}(\mu)$ can be determined from the large $k$ asymptotic behavior
of the zeros of the Airy function $\omega_k=(3\pi k/2)^{2/3}
+{\cal O}(k^{-1/3})$ to give ${\cal I}(\mu)\sim (2\sqrt{\pi}
(a\mu)^{3/2})^{-1}$. One thus get
\beq
\rho_1(\mu)=\sqrt{{a^3\over\pi\mu}}+{\cal O}(\mu^{1/2}),\quad \mu\to 0.
\label{as1}
\eeq
The divergence $\mu^{-1/2}$, as $\mu\to 0$, 
has been found in \cite{Ave-E} and 
seen in numerical simulations \cite{Kida}.

On the other hand, for large $\mu$, one can estimate the behavior of the 
function ${\cal H}(\mu)$ by the Laplace method to find
${\cal H}(\mu)\sim (\pi a^3 \mu^3)^{1/2}\exp(-a^3\mu^3/12)$. The behavior 
of the function ${\cal I}(\mu)$ is immediately given by the largest
zero of the Airy function to give ${\cal I}(\mu)\sim 
\exp(-\omega_1 a\mu)$. We thus have 
\beq
\rho_1(\mu)=2\sqrt{\pi}a^{9/2}\mu^{5/2}\exp\left(-{a^3\mu^3\over 12}
-\omega_1 a\mu\right),\quad\mu\to\infty.
\label{as2}
\eeq

\begin{figure}
\epsfxsize=11truecm
\hspace{2.75truecm}
\epsfbox{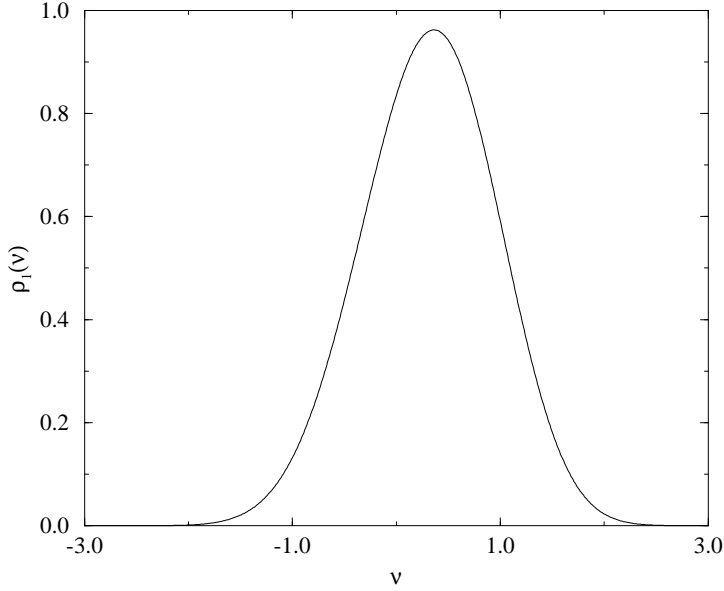}
\caption{Shock wavelength distribution $\rho_1(\nu)$ for $D=1/2$, 
($a=1$ in Eq.(\ref{pm})).
\label{fig5}}
\end{figure}

Let us consider now the shocks wavelength distribution. 
The one shock distribution (\ref{distrib})
can be written for the strength-wavelength variables $(\mu,\nu)$
as\footnote{The additional $\mu$ factor is the Jacobian of the 
transformation $(\mu,\eta)$ to $(\mu,\nu)$.}
\beq
\rho_1(\mu,\nu)=2a^4\mu{\cal J}(-\nu)
{\cal I}(\mu){\cal J}(\nu-\mu).
\eeq
Considering the variable $\nu'=\nu-\mu/2$ we find that the $(\mu,\nu')$ 
distribution is symmetric in $\nu'$, implying $\lla \nu'\rra=0$ and thus
$\lla\nu\rra={\lla\mu\rra\over 2}={1\over 2}$.   
The wavelength distribution $\rho_1(\nu)=\int_0^\infty d\mu\, \rho_1(\mu,\nu)$
is plotted on Fig. \ref{fig5}. Its asymptotic behaviour is found to be
$\rho_1(\nu)\sim C_{+}\nu^{3}\exp\(-a^3\nu^{3}/3-a\nu \omega_{1}\)$,
$\nu\to\infty$, and $\rho_1(\nu)\sim C_{-}\exp\(a^3\nu^3/3+a\nu \omega_{1}\)$,
$\nu\to -\infty$.
Remark that the wavelength 
distribution is not symmetrical around $\nu=\frac{1}{2}$.

\begin{figure}
\epsfxsize=11truecm
\hspace{2.75truecm}
\epsfbox{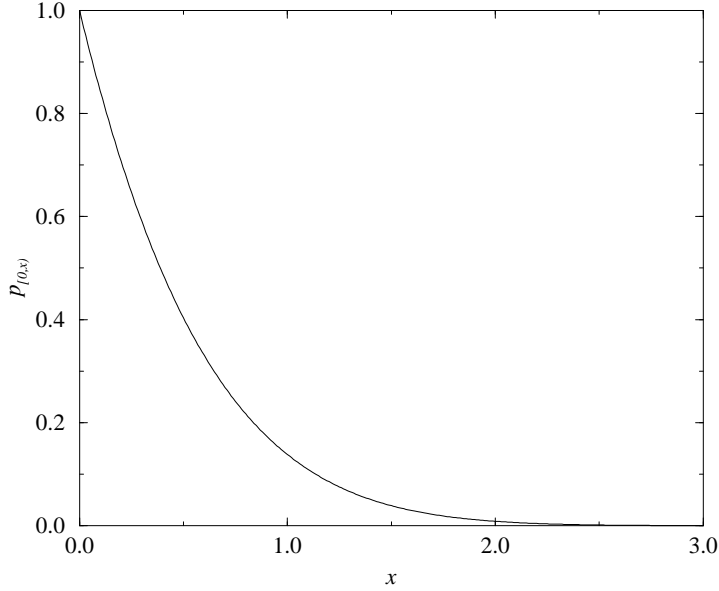}
\caption{
Distribution $p_{[0,x)}$ of intervals $[0,x)$ 
which contains no shocks for $D=1/2$
($a=1$ in Eq.(\ref{noshocks})).
\label{fig6}}
\end{figure}

The density distribution $p_{[0,x)}$ of intervals of size $x$ with no shocks 
(\ref{2.11e}) is given by 
\bea
p_{[0,x)}&=& \int_{-\infty}^{\infty}du_1\,J(-u_1)J(x+u_1)\non
&=& \sqrt{{\pi\over ax}}
\exp\(-{a^3x^3\over 12}\){1\over (2\pi\,i)^2}\int_{-i\infty}^{i\infty}
d\omega_1\int_{-i\infty}^{i\infty}d\omega_2\,
{\exp\({ax\over 2}(\omega_1+\omega_2)
+{(\omega_1-\omega_2)^2\over 4ax}\)\over \Ai(\omega_1)\Ai(\omega_2)}
\label{noshocks}
\eea
which is plotted on Fig. \ref{fig6}. 
Since $\lim_{x\to 0}p_{[0,x)}(u_1)=p_1(u_1)$, (see Eq.(\ref{2.11d})), 
and $p_1(u)$ is normalized (\ref{7f}), we have $\lim_{x\to 0} p_{[0,x)}=1$. 
Asymptotically we have for $x\to \infty$
\beq
p_{[0,x)}\sim\sqrt{{\pi\over ax}}{\exp\(-{a^3x^3\over 12}-a\omega_1 x\)\over
\left[\Ai'(-\omega_1)\right]^2}\(1+{\cal O}\({1\over x}\)\).
\eeq

\section{Correlations}
\label{section5}

In this section we study the two-point distributions of the Burgers 
velocity field and of the shocks in the asymptotic limit 
$x\to\infty$, keeping all the other
arguments fixed. From (\ref{2.11c}) and (\ref{2.15}), we have for $x$
large enough
\beq
p_{2}(x,u_{1},u_{2})=J(-u_{1})H(x,u_1,u_2)J(u_2)
\label{5.1}
\eeq
and 
\beq
\rho_2(x,\mu_{1},\eta_{1},\mu_{2},\eta_{2})=J(-\nu_{1})I(\mu_{1},\eta_{1})
H(x,-\mu_1+\nu_1,\nu_2)I(\mu_2,\eta_2)J(-\mu_2+\nu_2)
\label{5.2}
\eeq
with the function $J$ and $I$ given by Eqs.(\ref{jjj},\ref{iii}) and where the
function $H$ is defined by 
Eq.(\ref{2.11}) with $\nu_i=\mu_i/2-\eta_i/\mu_i$. 

Our main results are 
\bea
&&p_{2}(u_{1},u_{2},x)-p_{1}(u_{1})p_{1}(u_{2})\nonumber\\
&&\quad\quad \sim-
{8\sqrt{\pi}\over a^{1/2}x^{5/2}}\exp\(-\frac{a^3x^{3}}{12}-a\omega_{1}x\)
\exp\(-a\omega_{1}(u_{1}-u_{2})\){\cal J}(-u_1){\cal J}(u_2),\quad x\to\infty
\label{c0}
\eea
and similarly for the distribution of shocks 
\bea
\rho_{2}(\mu_{1},\eta_{1},&& \mu_{2},\eta_{2},x)
-\rho_{1}(\mu_{1},\eta_{1})\rho_{1}(\mu_{1},\eta_{1})
\sim -a^{11/2}{32\sqrt{\pi}\over x^{5/2}}
\exp\(-\frac{a^3x^{3}}{12}-a\omega_{1}x\)\non
&& \times \exp\(-a\omega_{1}(\nu_1-\nu_2-\mu_1)\)
{\cal J}(-\nu_1){\cal I}(\mu_1){\cal I}(\mu_2){\cal J}(-\mu_2+\nu_2)
,\quad x\to\infty.
\label{c0a}
\eea
Wee see that long distance correlations are very weak since they are again
dominated by the cubic decaying factor $\exp(-a^3x^3/12)$.

Clearly, in view of (\ref{5.1}) and (\ref{5.2}), 
this asymptotic behavior is determined by  
that of the function $H(x,\nu_{1},\nu_{2})$.
First we write $H(x,\nu_{1},\nu_{2})$ in explicit form by 
introducing (\ref{12}) in (\ref{2.11}). 
It is useful to remember that by the definition 
of $(\mu^{*},\eta^{*})$ one has 
$\eta^{*}=s_{\nu_{1}}(\mu^{*})=q+s_{\nu_{2}}(\mu^{*}-x-\nu_{1}+\nu_{2})$.
To bring the expression in the 
most symmetric form the change of integration variables
\bea
\zeta&=&(D^{2}/2)^{1/3}(\eta^{*}-\eta^{\prime}),\;\;\;\;\;\;\;\;
0< \zeta<\infty\nonumber\\
r &=&\frac{1}{\sqrt{x}}\(q+\frac{\nu_{1}^{2}}{2}-\frac{\nu_{2}^{2}}{2}\),\;\;
-\srx r_{1}\leq r\leq \srx r_{2},\;\;\;
r_{1}=\frac{x}{2}+\nu_{1},\;\;\;r_{2}=\frac{x}{2}-\nu_{2}
\label{c0b}
\eea
turns out to be adequate. Then, with $a=(2D)^{-1/3}$, one obtains 
\bea
&& H(x,\nu_{1},\nu_{2})=2a^{3}
\exp\(\frac{-a^{3}\nu_{1}^{3}+a^{3}\nu_{2}^{3}}{3}\)\srx
\exp\(-{a^3x^{3}\over 12}\)
\int_{-\srx r_{1}}^{\srx r_{2}} dr\exp\(-a^{3}r^{2}\)\non
&&\quad \times \int _{0}^{\infty}d\zeta e^{a\zeta x}
\sum_{k_{1},k_{2}}\exp\left[-a\omega_{k_{1}}\(r_{1}+\frac{r}{\srx}\)
-a\omega_{k_{2}}\(r_{2}-\frac{r}{\srx}\)\right]
\frac{\Ai(\zeta-\omega_{k_{1}})
\Ai(\zeta-\omega_{k_{2}})}
{{\rm Ai}^{\prime}(-\omega_{k_{1}}) {\rm Ai}^{\prime}(-\omega_{k_{2}})}.
\label{c1}
\eea
Our main concern is to determine the asymptotic behavior 
of this expression as $x\to\infty$. 
We give here the main steps of the calculation 
while details and justifications are given in the appendices.

To get the basic clustering properties of the model, 
we expect that $\lim_{x\to\infty}H(x,\nu_{1},\nu_{2})=J(\nu_{1})J(-\nu_{2})$
with $J(\nu)$ given by the integral in the complex plane Eq.(\ref{jjj}). 
It is therefore natural to replace the sums on the zeros of 
Airy functions in (\ref{c1}) 
by appropriate contour integrals, as in Sec.\ref{section4},
\bea
&&\sum_{k_{1},k_{2}}\exp\left[-a\omega_{k_{1}}\(r_{1}+\frac{r}{\srx}\)
-a\omega_{k_{2}}\(r_{2}-\frac{r}{\srx}\)\right]
\frac{\Ai(\zeta-\omega_{k_{1}})\Ai(\zeta-\omega_{k_{2}})}
{{\rm Ai}^{\prime}(-\omega_{k_{1}}) {\rm Ai}^{\prime}(-\omega_{k_{2}})}=\non
&&\quad {1\over (2\pi i)^2}\int_{{\cal C}_{x}}dw_{1}\int_{{\cal C}_{x}}dw_{2}
\exp\left[aw_1\(r_{1}+\frac{r}{\srx}\)
+aw_2\(r_{2}-\frac{r}{\srx}\)\right]
\frac{\Ai(\zeta+w_{1})\Ai(\zeta+w_{2})}{\Ai(w_{1})\Ai(w_{2})}
\label{c2a}
\eea
For  a given $x$ the contour ${\cal C}_{x}$ 
is chosen as the parabola with branches 
$w^{\pm}(\rho)=-\rho\pm i\,ax\sqrt{\rho}, \;0\leq \rho <\infty$. 
This contour will be convenient to
determine the large $x$ asymptotics of $H(x,\nu_{1},\nu_{2})$. 
The integrals (\ref{c2a}) on ${\cal C}_{x}$  converge for $r$ 
fixed because of the exponentially decreasing factors 
$\exp\left[aw_1\(r_{1}+\frac{r}{\srx}\)+aw_2\(r_{2}-\frac{r}{\srx}\)\right], 
\Re w_{1}<0, \;\Re w_{2}<0,\;
r_{1}+\frac{r}{\srx}>0,\; r_{2}-\frac{r}{\srx}>0$ (see Appendix \ref{app1}).

Next we exchange the $\zeta$-integral with
the contour integrals to obtain
\bea
&& H(x,\nu_{1},\nu_{2})=
2a^{3}
\exp\(\frac{-a^{3}\nu_{1}^{3}+a^{3}\nu_{2}^{3}}{3}\)\srx
\exp\(-\frac{a^3 x^{3}}{12}\)
\int_{-\srx r_{1}}^{\srx r_{2}} dr\exp\(-a^{3}r^{2}\)\non
&&
\quad\quad \times
{1\over (2\pi i)^2}\int_{{\cal C}_{x}}dw_{1}\int_{{\cal C}_{x}}dw_{2}
\exp\left[aw_1\(r_{1}+\frac{r}{\srx}\)
+aw_2\(r_{2}-\frac{r}{\srx}\)\right]
\frac{B(ax,w_1,w_2)}{\Ai(w_{1})\Ai(w_{2})}
\label{c2b}
\eea
where
\beq
B(x,w_{1},w_{2})=\int_{0}^{\infty}d\zeta e^{\zeta x}\Ai(\zeta+w_{1})
\Ai(\zeta+w_{2})
\label{c3}
\eeq
is the Laplace transform of a product of Airy functions 
evaluated at the negative argument $-x$.
This Laplace transform is computed in Appendix \ref{app2} and
is given as the difference of two terms $B(x,w_{1},w_{2})
=B_{1}(x,w_{1},w_{2})-B_{2}(x,w_{1},w_{2})$ (see Eq.(\ref{L5})).
We set $H(x)=H_{1}(x)-H_{2}(x)$ with $H_{1}(x)$ 
(resp. $H_{2}(x)$) the contribution to (\ref{c2b}) 
of $B_{1}(x,w_{1},w_{2})$
(resp. $B_{2}(x,w_{1},w_{2})$). 
Then
\beq
H_{1}(x)=
\frac{a^{5/2}}{\sqrt{\pi}}
\exp\(\frac{-a^{3}\nu_{1}^{3}+a^{3}\nu_{2}^{3}}{3}\)
\int_{-\srx r_{1}}^{\srx r_{2}} dr
{1\over (2\pi i)^2}\int_{{\cal C}_{x}}dw_{1}
\int_{{\cal C}_{x}}dw_{2}\,
h_{1}(\eta,w_{1},w_{2})
\label{c4a}
\eeq
with
\beq
h_{1}(r,w_{1},w_{2})=
{\exp\left[-a^3\(r-{w_1-w_2\over 2a^2\srx}\)^2+aw_1\nu_1-aw_2\nu_2\right]
\over
\Ai(w_{1})\Ai(w_{2})}.
\label{c5}
\eeq
It is shown in appendix \ref{app3} 
that the multiple integral in (\ref{c4a}) is absolutely convergent. 
As $x\to\infty$, the
contour ${\cal C}_{x}$ eventually opens to the imaginary axis 
of the $w$-plane. Hence one sees (formally) on (\ref{c4a}) that
\bea
\lim_{x\to\infty}H_{1}(x)&=&
a\exp\(\frac{-a^3\nu_1^{3}+a^{3}\nu_2^3}{3}\)
{1\over (2\pi i)^2}\int_{-i\infty}^{i\infty}dw_{1}
\int_{-i\infty}^{i\infty}dw_{2}\frac
{\exp\(aw_1\nu_1-aw_2\nu_2\)}
{\Ai(w_{1})\Ai(w_{2})}\non
&=& J(\nu_1)J(-\nu_2)
\label{c6}
\eea 
where the function $J(\nu)$ is defined by Eq.(\ref{jjj}).
More precisely one finds that the asymptotic behavior 
of $H_{1}(x)$ is given by (Appendix \ref{app3})
\beq
H_{1}(x)=J(\nu_1)J(-\nu_2)+{\cal O}\(\exp\(-\frac{a^3x^{3}}{12}(1+c)\)\)
, \;\;\,c>0.
\label{c7}
\eeq

Inserting the expression $B_{2}(x)$  (\ref{L5}) in (\ref{c2b}) one finds 
\bea
&&H_{2}(x)=2a^{5/2}\exp\(\frac{-a^3\nu_1^{3}+a^3\nu_2^{3}}{3}\)
\int_{-\srx r_{1}}^{\srx r_{2}} dr\,{\rm e}^{-a^3r^2}
\non
&&\quad \times
{1\over (2\pi i)^2}\int_{{\cal C}_{x}}dw_{1}\int_{{\cal C}_{x}}dw_{2}
\exp\(aw_{1}\(r_{1}+\frac{r}{\srx}\)+aw_{2}\(r_{2}-\frac{r}{\srx}\)\)
\int_{ax}^{\infty}dy\sqrt{y}
\non
&&\quad
\times\exp\(-\frac{y^{3}}{12}+{w_{1}+w_{2}\over 2}(y-ax)
-{(w_{1}-w_{2})^{2}\over 4}\(\frac{1}{ax}-\frac{1}{y}\)\)
\frac{g(y,w_{1},w_{2})}{\Ai(w_{1})\Ai(w_{2})}.
\label{c7a}
\eea
Because of the convergence factors $\exp\(aw_{1}\(r_{1}+\frac{r}{\srx}\)
+aw_{2}\(r_{2}-\frac{r}{\srx}\)\)$
the contours ${\cal C}_{x}$ can be closed 
and the corresponding integrals can again be 
evaluated at the zeros of the Airy functions (the arguments
are similar to those given in appendix \ref{app1}). 
Then the relation (\ref{L6}) permits to simplify the result to
\bea
&& H_{2}(x)=2a^{5/2}\exp\(\frac{-a^3\nu_1^{3}+a^3\nu_2^{3}}{3}\)
\int_{-\srx r_{1}}^{\srx r_{2}} dr\,{\rm e}^{-a^3r^2}\non
&&\quad\times \sum_{k_1,k_2}\exp\(-a\omega_{k_1}\(r_{1}
+\frac{r}{\srx}\)-a\omega_{k_2}
\(r_{2}-\frac{r}{\srx}\)\)\non
&&\quad\times\int_{ax}^{\infty}dy{1\over\sqrt{y}}
\exp\(-\frac{y^{3}}{12}-{\omega_{k_1}+\omega_{k_2}\over 2}
(y-ax)
-{(\omega_{k_1}-\omega_{k_2})^{2}\over 4}\(\frac{1}{ax}-\frac{1}{y}\)\)
\label{c8}
\eea
To compute the large $x$ behavior it is convenient 
to make the change of integration variable
$y=\frac{z}{x^{2}}+ax$ giving
\beq
H_2(x)=2a^{2}\exp\(\frac{-a^3\nu_1^{3}+a^3\nu_2^{3}}{3}\)
\frac{\exp\(-\frac{a^3x^{3}}{12}\)}{x^{5/2}}G(x)
\label{c8a}
\eeq
with
\bea
G(x)=\int_{0}^{\infty}dz &&\frac{\exp\(-\frac{z^{3}}{12x^{6}}
-\frac{az^{2}}{4x^{3}}-\frac{a^2z}{4}\)}{\sqrt{1+\frac{z}{ax^{3}}}}
\int_{-\srx r_{1}}^{\srx r_{2}} dr\,{\rm e}^{-a^3r^2}
\sum_{k_{1},k_{2}}
\exp\(-(\omega_{k_{1}}-\omega_{k_{2}})^{2}\frac{z}{4a(zx+ax^{4})}\)
\non
&&\quad\quad\times
\exp\(-a\omega_{k_{1}}\(r_{1}+\frac{r}{\srx}+\frac{z}{2ax^{2}}\)
-a\omega_{k_{2}}\(r_{2}-\frac{r}{\srx}+\frac{z}{2ax^{2}}\)\)
\label{c9}
\eea
Letting formally $x\to\infty$ on this formula 
gives the asymptotic
behavior (details are found in appendix \ref{app4})
\beq
H_{2}(x)\sim {8\sqrt{\pi}\over a^{3/2}x^{5/2}}
\exp\(\frac{-a^3\nu_1^{3}+a^3\nu_2^{3}}{3}-a\omega_{1}(\nu_1-\nu_2)\)
\exp\(-\frac{a^3x^{3}}{12}-a\omega_{1}x\)
\label{c10}
\eeq
where $-\omega_{1}$ is the first zero of the Airy function.

Inserting the asymptotics Eqs.(\ref{c10},\ref{c7}) in the expression for the
two-point distributions Eqs.(\ref{5.1},\ref{5.2}) leads to the 
results Eqs.(\ref{c0},\ref{c0a}).

\section{Conclusion}

To conclude, we remark that the previous results allow for a complete
statistical description of the Burgers field.
As mentioned in the introduction, for white noise initial data, 
$u(x)$ is a Markov process as function of $x$ \cite{Ave-E}. Thus
with $P(x_{2},u_{2}|x_1,u_1)=p_2(x_{2}-x_1,u_1,u_{2})/p_1(u_1)$ the transition
kernel for the Markov process, the $n$-point distribution can be written
as
\bea
p_n(x_1,u_1;\ldots ;x_n,u_n) &=& P(x_n,u_n|x_{n-1},u_{n-1})\ldots 
P(x_2,u_2|x_1,u_1)p_1(x_1,u_1)\non
&=& {\prod_{i=1}^{n-1}p_2(x_{i+1}-x_i,u_i,u_{i+1})
\over \prod_{i=2}^{n-1} p_1(u_i)},\quad n\geq 3.
\eea
On the same line, a complete statistical description of shocks in Burgers 
solution is obtained through the $n$-shocks distribution densities which 
factorize to
\beq
\rho_n(x_{1},\mu_{1},\eta_{1},\ldots,x_{n},\mu_{n},\eta_{n})=
{\prod_{i=1}^{n-1}
\rho_2(x_{i+1}-x_i,\mu_i,\eta_i;\mu_{i+1},\eta_{i+1})\over
\prod_{i=2}^{n-1}\rho_1(\mu_i,\eta_i)},\quad n\geq 3.
\label{cc2}
\eeq
The distribution of ordered sequences of next neighboring shocks is obtained
from (\ref{cc2}) by omitting the function $H$ in $\rho_2$, Eq.(\ref{2.15}).
Here, factorization follows simply from the Markov property of Brownian motion
and the fact that multiple constraints of the form (\ref{7d}) decouple.
From  the point of view of the hierarchy of kinetic equations that
governs the dynamics of shocks this factorization corresponds to an 
exact closure of this hierarchy or to an exact propagation of chaos.
This will be discussed in \cite{Fra-Mar-Pia}. 

As far as the time dependence is concerned, 
it can be reintroduced via the basic 
transition kernel (\ref{12}), which should be computed with $s_\nu(y)$ 
replaced by $s_\nu(y)/t$. Owing to the invariance of the Brownian 
measure under the change $\psi(y)\to t^{1/3}\psi(y/t^{2/3})$, one
immediately finds that
$K_\nu(\mu_1,\eta_1,\mu_2,\eta_2;t)=t^{-1/3}
K_{\nu^{\prime}}(\mu_1^{\prime},\eta_1^{\prime},\mu_2^{\prime}
,\eta_2^{\prime})$
where the variables are rescaled according to
$\mu_i^{\prime}=\mu_i t^{-2/3}$, $\eta_i^\prime=\eta_i t^{-1/3}$, and
$\nu^{\prime}=\nu t^{-2/3}$.
From (\ref{2.8}), (\ref{2.14}) and (\ref{2.11}) this implies the
transformation laws of the functions $J$, $I$, and $H$
\beq
J(\nu;t)=t^{-1/3}J(\nu^{\prime}),\quad 
I(\mu,\eta;t)=t^{-1}I(\mu^{\prime},\eta^{\prime}),
\;{\rm and}\quad H(x,\nu_1,\nu_2,t)=t^{-2/3}
H(x^{\prime},\nu_1^{\prime},\nu_2^{\prime})
\eeq
where $x^{\prime}=xt^{-2/3}$.
This leads to the time dependent distributions
\beq
p_n(x_1,u_1;\ldots ;x_n,u_n;t)=t^{n/3} p_n(x_1^{\prime},u_1^{\prime};\ldots;
x_n^{\prime},u_n^{\prime}),
\label{102}
\eeq
with $u_i^\prime=u_i t^{1/3}$, and 
\beq
\rho_n(x_{1},\mu_{1},\eta_{1};\ldots;x_{n},\mu_{n},\eta_{n};t)=
t^{-5n/3}\rho_n(x_{1}^{\prime},\mu_{1}^{\prime},\eta_{1}^{\prime};\ldots;
x_{n}^{\prime},\mu_{n}^{\prime},\eta_{n}^{\prime}).
\eeq
To obtain (\ref{102}), we recall that the distributions 
$p_n(x_1,u_1;\ldots ;x_n,u_n)$ were calculated from those of the
coordinates of the contact points $x_i-\xi_i$. At time $t\ne 1$, one has 
$x_i-\xi_i=u_it$ introducing a Jacobian $t^n$ included in (\ref{102}) when
expressing the distributions as functions of the Burgers field amplitudes 
$u_i$. From there, one recovers 
the well-known time dependent behavior of 
some moments of the distributions {\sl e.g.}, 
the energy dissipation per unit of length 
$\lla u^2(x,t)\rra\sim t^{-2/3}$, the average number of shocks per unit 
of length $\sim t^{-2/3}$, the average strength of a shock $\lla \mu/t\rra
\sim t^{-1/3}$.

\acknowledgements

We thank J. Piasecki for many useful discussions.

\appendix
\section{}
\label{app1}

We justify in this appendix the equation (\ref{c2a}) which  replaces
the sum on zeros of the Airy function by an integral in the complex plane.

To evaluate $\Ai(w)$ on the branch 
$w^{+}(\rho)=-\rho+i\,ax\sqrt{\rho}$, 
we start from the formula
$\Ai(-w)=e^{i\pi/3}\Ai(we^{i\pi/3})+e^{-i\pi/3}\Ai(we^{-i\pi/3})$ 
\cite{Abra} 
giving\footnote{The formula enables
to obtain the asymptotic behavior of the Airy function  $\Ai(w)$ 
when $\arg w$ approaches $\pi$ as it is the case for  
$w^{\pm}(\rho)$, $\rho\to\infty$.}
\bea
\Ai(w^{+}(\rho))&=&e^{i\pi/3}\Ai(-w^+(\rho)e^{i\pi/3})
+e^{-i\pi/3}\Ai(-w^+(\rho)e^{-i\pi/3})\nonumber\\
&\sim&\(\frac{1}{\sqrt{4\pi}(w^+(\rho))^{1/4}}\)
\(\exp\(-i\,\frac{2}{3}(-w^+(\rho))^{3/2}\)
+\exp\(i\,\frac{2}{3}(-w^+(\rho))^{3/2}\)\)
\label{O1}
\eea
where we have used the asymptotic behavior (\ref{jj3}) of the Airy 
function $\Ai(w)$ for $|w|\to\infty$, ${\rm arg}\,w\neq \pi$.
As $\rho\to\infty$, 
\bea
(-w^+(\rho))^{3/2}&=&\(\rho-i\,ax\sqrt{\rho}\)^{3/2}
=\rho^{3/2}-i\,\frac{3}{2}ax\rho-\frac{3}{8}a^2x^{2}\sqrt{\rho}
-i\,a^3\frac{x^{3}}{16}
+{\rm O}\(\frac{x^{4}}{\sqrt{\rho}}\)
\label{O2}
\eea
Upon inserting (\ref{O2}) in (\ref{O1}) one sees that 
\beq
\left|\frac{1}{\Ai(w^{+}(\rho))}\right|\leq C_{\rho,x}
\exp\(-ax\rho-\frac{a^3x^{3}}{24}\)
\label{O3}
\eeq
with $C_{\rho,x}$ growing at most algebraically with $\rho$ and $x$.
Using $\Ai(w^{*})=\Ai^{*}(w)$ one has the same estimate 
on the branch $w^{-}(\rho)$.
By a similar calculation one has also that, for
fixed $\zeta$, $\Ai(\zeta+w^{\pm}(\rho))/\Ai(w^{\pm}(\rho)$  
remains bounded  as $\rho\to\infty$.

Consider now the finite parabolic contour closed by a 
circular arc ${\cal R}e^{i\theta}$ with $\theta$ close to $\pi$. On
this circular arc for large radius ${\cal R}$ 
\bea
\frac{\Ai(\zeta+{\cal R}e^{i\theta})}{\Ai({\cal R}e^{i\theta})}&\sim& 
\(\frac{{\cal R}e^{i\theta}}{\zeta+{\cal R}e^{i\theta}}\)^{1/4}
\exp\(-\frac{2}{3}(\zeta+{\cal R}e^{i\theta})^{3/2}+\frac{2}{3}
({\cal R}e^{i\theta})^{3/2}\)\non
&\sim&\(\frac{{\cal R}e^{i\theta}}{\zeta+{\cal R}e^{i\theta}}\)^{1/4}
\exp\(-\zeta\sqrt{{\cal R}}e^{i\theta/2}\)={\cal O}(1)
\label{O4}
\eea
as ${\cal R}\to\infty$ and $\frac{\pi}{2}\leq \theta \leq \pi$. Since
$\(r_{1}+\frac{r}{\srx}\)>0,\;\(r_{2}-\frac{r}{\srx}\)>0$ 
the factors
$\exp\(aw_{1}\(r_{1}+\frac{r}{\srx}\)\)$ and $\exp\(aw_{2}
\(r_{2}-\frac{r}{\srx}\)\)$
decay exponentially fast when $w_{1}$ and $w_{2}$ are on the 
contour ${\cal C}_{x}$ 
or on the circular arc. One concludes that the integrals 
on the circular arcs vanish as ${\cal R}\to\infty$
so that the sums in (\ref{c2a}) can indeed be replaced 
by the contour integrals.

\section{}
\label{app2}

The integral $B(x)$ (\ref{c3}) 
\beq
B(x)=\int_{0}^{\infty}d\zeta e^{\zeta x}\Ai(\zeta+w_{1})
\Ai(\zeta+w_{2})
\eeq
is the Laplace transform for a negative argument $-x$
of the product $f(\zeta)=\Ai(\zeta+w_{1})\Ai(\zeta+w_{2})$ of two
Airy functions (omitting $w_{1}$ and $w_{2}$ from the notation). 
First the asymptotic behavior 
of $B(x)$ is determined
by the Laplace method 
\beq
B(x) \sim \frac{1}{2\sqrt{\pi}}e^{\Phi(x)},\;\;\;x\to\infty
\label{L1}
\eeq
where
\beq
\Phi(x)={x^3\over 12}-{x\over 2}(w_1+w_2)
-\frac{1}{2}\ln x-{(w_1-w_2)^2\over 4x}.
\label{L1b}
\eeq
From the property of the Airy function (\ref{14}), 
$f(\zeta)$ verifies the 4th order differential equation
\beq
f''''(\zeta)-(4\zeta+2w_1+2w_2)f''(\zeta)-6f'(\zeta)+(w_1-w_2)^2f(\zeta)=0.
\label{L2}
\eeq
From (\ref{L2}), one finds that its Laplace transform 
for negative arguments satisfies
\beq
B^{\prime}(x)-h(x)B(x)=g(x)
\label{L3}
\eeq
where we remark that
\beq
h(x)=\Phi^\prime(x)
\eeq
and with
\bea
g(x)=\frac{x}{4}f(0)-\frac{1}{4}f'(0)&+&\frac{1}{4x}\left[
(f''(0)-2(w_1+w_2)f(0)\right]\non
&-&\frac{1}{4x^{2}}\left[f'''(0)-2f(0)-2(w_1+w_2)f'(0)\right].
\label{L4}
\eea

Eq. (\ref{L3}) can be solved, 
using also the value (\ref{L1}) for $x\to\infty$, 
\beq
B(x)=B_1(x)-B_2(x)
\eeq
with
\beq
B_1(x)=\frac{1}{2\sqrt{\pi}}e^{\Phi(x)},\quad
B_2(x)=e^{\Phi(x)}\int_{x}^{\infty}dy e^{-\Phi(y)}g(y).
\label{L5}
\eeq
Notice that when evaluated at the zeros of the Airy functions 
$w_{1}=-\omega_{k_{1}},\;w_{2}=-\omega_{k_{2}}$,
$g(y)$ reduces to
\beq
g(y)\mid_{w_{1}=-\omega_{k_{1}},\;w_{2}=-\omega_{k_{2}}}
=\frac{\Ai^{\prime}(-\omega_{k_{1}})
\Ai^{\prime}(-\omega_{k_{2}})}{2y}.
\label{L6}
\eeq

\section{}
\label{app3}

We consider the multiple integral $H_1(x)$ (\ref{c4a}) and 
show first that it is absolutely convergent. On the contour
$w^\pm(\rho)=-\rho\pm i\,ax\sqrt{\rho}$, $0\leq \rho<\infty$, we have
\beq
\Re\(r-\frac{w_{1}-w_{2}}{2a^2\srx}\)^{2}=
\(r+\frac{\rho_{1}-\rho_{2}}{2a^2\srx}\)^{2}
-\frac{x}{4a^2}(\sqrt{\rho_{1}}\pm\sqrt{\rho_{2}})^{2}
\label{A2}
\eeq
Hence, using $(\sqrt{\rho_{1}}\pm\sqrt{\rho_{2}})^{2}
\leq 2(\rho_{1}+\rho_{2})$ and (\ref{O3}), 
the integrand (\ref{c5}) is bounded by    
\bea
&& \left| h_1(r,w_1,w_2)\right| \leq
C_{\rho_{1},\rho_{2},x}\exp\(-\frac{a^3x^{3}}{12}\)\times\non
&& \quad\quad \times\exp\left\{-ax(\rho_{1}+\rho_{2})-a\rho_1\nu_1+a\rho_2\nu_2
-a^3\(r+{\rho_1-\rho_2\over 2a^2\sqrt{x}}\)^2+{ax\over 4}\(\sqrt{\rho_1}\pm
\sqrt{\rho_2}\)^2\right\}\non
&&\quad\quad\quad\quad \quad\quad
\leq C_{\rho_{1},\rho_{2},x}\exp\(-\frac{a^3x^{3}}{12}\)
\exp\left\{-a\rho_1r_1-a\rho_2r_2
-a^3\(r+{\rho_1-\rho_2\over 2a^2\sqrt{x}}\)^2\right\}   
\label{A3}   
\eea
with $C_{\rho_{1},\rho_{2},x}$ increasing at most algebraically, 
showing that the integral (\ref{c4a}) converges absolutely.

To obtain the asymptotic behavior (\ref{c7}) of $H_{1}(x)$ we write the
integration of $h_1(r,w_1,w_2)$ over $r$ as 
\beq
\int_{-\srx r_1}^{\srx r_{2}} dr\, h_1(r,w_1,w_2)=
\left(\int_{-\infty}^\infty dr-\int_{-\infty}^{-\srx r_1}dr
-\int_{\srx r_2}^\infty dr\)h_1(r,w_1,w_2).
\eeq
The first integration is readily performed to give $J(\nu_1)J(-\nu_2)$ 
(see Eq.(\ref{c6})) as 
$J(\nu)$ (\ref{jjj}) can be represented as an integral 
on any contour that encircles the zeros of the Airy function,
in particular on  ${\cal C}_{x}$. 
Thus it follows from (\ref{c4a}) that
\bea
 H_{1}(x)-J(\nu_1)&& J(-\nu_2)=-\frac{a^{5/2}}{\sqrt{\pi}}
\exp\(\frac{-a^{3}\nu_1^{3}+a^{3}\nu_2^{3}}{3}\)\non
&& \times
\(\int_{-\infty}^{-\srx r_{1}}dr+\int_{\srx r_{2}}^\infty dr\) 
{1\over (2\pi i)^{2}}\int_{{\cal C}_{x}}dw_{1}
\int_{{\cal C}_{x}}dw_{2}\,
h_{1}(\eta,w_{1},w_{2})
\label{A4}
\eea
Consider the contribution to (\ref{A4}) 
where $r\geq\srx r_{2}$ and the branches of ${\cal C}_{x}$ are
$w^{+}(\rho_{1}),w^{+}(\rho_{2})$. 
With (\ref{A3}) this contribution is majorized by 
\bea
&&\left|\int_{\srx r_2}^\infty dr \int_{w^+} dw_1\int_{w^+} dw_2\, 
h(r,w_1,w_2)\right|\leq
\exp\(-\frac{a^3x^{3}}{12}\)\int_{\srx r_2}^\infty dr
\int_0^\infty d\rho_1 \int_0^\infty d\rho_2  \non
&&\quad\quad \times\left|\frac{dw^{+}(\rho_{1})}{d\rho_{1}}\right|
\left|\frac{dw^{+}(\rho_{2})}{d\rho_{2}}\right|
C_{\rho_{1},\rho_{2},x}\exp\left\{-a\rho_1r_1-a\rho_2r_2
-a^3\(r+{\rho_1-\rho_2\over 2a^2\sqrt{x}}\)^2\right\} 
\label{A5}
\eea
We split the $\rho_{2}$ integral into the 
domains $0\leq \rho_2 \leq a^2\srx r$  and $a^2\srx r\leq \rho_2 <\infty$.
When $0\leq \rho_2\leq a^2\srx r$, $\rho_1\geq 0$, 
$r\geq \srx r_{2}=\srx(x/2-\nu_2)$ 
one has 
\beq
\(r+{\rho_1-\rho_2\over 2a^2\srx}\)^2
\geq \({r\over 2}+{\rho_1\over 2a^2\srx}\)^2
\geq \({r\over 2}\)^2\geq {r^2\over 8}+{r_2^2 x\over 8}\geq 
{r^2\over 8}+cx^3
\eeq
where the last inequality holds for $x$ large enough with $c>0$, and thus
\beq
\exp\left\{-a^3\(r+{\rho_1-\rho_2\over 2a^2\sqrt{x}}\)^2\right\}\leq
\exp\left\{-a^3 {r^2\over 8}-ca^3x^3\right\}.
\label{A6}
\eeq
On the other hand, when $\rho_2\geq a^2\srx r$, $r\geq \srx r_{2}$,
\beq
\rho_2r_2\geq {\rho_2r_2\over 2}+{a^2r\srx  r_2\over 2}\geq 
{\rho_2r_2\over 2}+{a^2 r_2^2\over 2}\geq {\rho_2r_2\over 2}+cx^3
\eeq
where the last inequality holds for $x$ large enough with $c>0$. This leads to
\beq
\exp\(-a\rho_2 r_2\)\leq \exp\(-{a\rho_2 r_2\over 2}-acx^3\).
\label{A7}
\eeq 
The bounds (\ref{A6}) and (\ref{A7}) are introduced in (\ref{A5}), 
the remaining $r,\;\rho_1,\;\rho_2$ integrals
are convergent and bounded with respect to $x$ 
(except for a polynomial growth due to $C_{\rho_1,\rho_{2},x}$ 
and the line elements 
$\mid\frac{dw(\rho)}{d\rho}\mid=\sqrt{1+\frac{a^2x^{2}}{4\rho}}$). 
The other contributions to (\ref{A4}) are treated in the same way.
This leads to the result (\ref{c7}).

\section{}
\label{app4}

We determine here the asymptotic behavior of $G(x)$ (\ref{c9}) for
large $x$. Starting from
\beq
G(x)=\int_{0}^{\infty}dz\int_{-\srx r_{1}}^{\srx r_{2}} dr 
\sum_{k_{1}\geq 1}\sum_{k_{2}\geq 1} {\cal G}_{k_1,k_2}(z,r;x)
\eeq
with
\bea
{\cal G}_{k_1,k_2}(z,r;x)= &&\frac{\exp\(-\frac{z^{3}}{12x^{6}}
-\frac{az^{2}}{4x^{3}}-\frac{a^2z}{4}\)}{\sqrt{1+\frac{z}{ax^{3}}}}
{\rm e}^{-a^3r^2}
\exp\(-(\omega_{k_{1}}-\omega_{k_{2}})^{2}\frac{z}{4a(zx+ax^{4})}\)\non
&&\quad\times
\exp\left[-a\omega_{k_{1}}\(r_{1}+\frac{r}{\srx}+\frac{z}{2ax^{2}}\)
-a\omega_{k_{2}}\(r_{2}-\frac{r}{\srx}+\frac{z}{2ax^{2}}\)\right]
\eea
we define
\beq
F(x)={\rm e}^{a\omega_{1}r_{1}+a\omega_{1}r_{2}}G(x)
\label{B1}
\eeq
where $r_1=x/2+\nu_1$, $r_2=x/2-\nu_2$ and $-\omega_1$ is the largest zero
of the Airy function. 
We then decompose $F(x)= F_{a}(x)+F_{b}(x)+F_{c}(x)$ 
according to the following splitting of the $r$ integration range
and the $k_{1},k_{2}$ summations (for $x$ large):
\bea
F_{a}(x)&=&\int_{0}^{\infty}dz\int_{-r_1}^{r_2}dr\,
{\rm e}^{a\omega_{1}r_{1}+a\omega_{1}r_{2}}
{\cal G}_{1,1}(z,r;x)\label{B2}\\ 
F_{b}(x)&=&\int_{0}^{\infty}dz\int_{-r_{1}}^{r_{2}}dr 
{\rm e}^{a\omega_{1}r_{1}+a\omega_{1}r_{2}}\(
\sum_{k_{1}\geq 1}\sum_{k_{2}\geq 1} {\cal G}_{k_1,k_2}(z,r;x)-
{\cal G}_{1,1}(z,r;x)\)
\label{B3}\\
F_{c}(x)&=&\int_{0}^{\infty}dz\(\int_{r_{2}}^{\srx r_{2}}dr
+\int_{-\srx r_{1}}^{-r_{1}}dr \)
{\rm e}^{a\omega_{1}r_{1}+a\omega_{1}r_{2}}
\sum_{k_{1}\geq 1}\sum_{k_2\geq 1}{\cal G}_{k_1,k_2}(z,r;x)
\label{B4}
\eea
By dominated convergence, we immediately have that
\beq
\lim_{x\to\infty}F_{a}(x)=
\int_0^\infty dz\,\exp\(-{a^2 z\over 4}\)
\int_{-\infty}^{\infty}dr\,\exp\(-a^3r^2\)
={4\sqrt{\pi}\over a^{7/2}}.
\label{B5}
\eeq
We then show below that $F_{b}(x)$ and $F_{c}(x)$ vanish as $x\to\infty$
leading to the asymptotic behavior 
\beq
G(x)\sim {4\sqrt{\pi}\over a^{7/2}}{\rm e}^{-a\omega_1(x+\nu_1-\nu_2)}
,\quad (x\to\infty)
\eeq
and thus to the behavior of $H_2(x)$, Eq.(\ref{c10}).

Since $-r_{1}\leq r\leq r_{2}$ in the integral (\ref{B3}), 
one can choose $x$ large enough so that
$r_{1}+\frac{r}{\srx}\geq r_{1}(1-\epsilon)$, 
$r_{2}-\frac{r}{\srx}\geq r_{2}(1-\epsilon)$, $\epsilon>0$.
Hence the $k_{1},k_{2}$ term of the integrand in (\ref{B3}) is less than
\beq
{\rm e}^{a\omega_{1}r_{1}+a\omega_{1}r_{2}}{\cal G}_{k_1,k_2}(z,r;x)\leq
{\rm e}^{-a^3r^2-a^2z/4}{\rm e}^{-ar_1(\omega_{k_1}(1-\epsilon)-\omega_1)}
{\rm e}^{-ar_2(\omega_{k_2}(1-\epsilon)-\omega_1)}
\label{B6}
\eeq
showing that the joint $z$, $r$ integrals and $k_{1}$, $k_{2}$ 
summations converge. Moreover, since the term $(k_1,k_2)=(1,1)$ is absent 
from the integrand in (\ref{B3}), 
there is at least one of the indices strictly greater than one.
If both the indices are strictly greater than one,
we can conclude that
$0<F_{b}(x)\leq C\exp(-a\min(r_{1},r_{2})(\omega_{2}(1-\epsilon)-\omega_{1}))$
tends to zero exponentially fast
as $x\to\infty$ provided that $\epsilon<(\omega_{2}-\omega_{1})/\omega_{2}$ 
with $-\omega_{2}$ 
the second zero of the Airy function. If one of the indices is equal to one, 
say $k_1=1$, $k_2>1$, we have
$0<F_{b}(x)\leq C\exp(-ar_2(\omega_{2}(1-\epsilon)-\omega_{1})+a\epsilon r_1)$
which tends exponentially to zero as $x\to \infty$ provided that
$\epsilon<(\omega_{2}-\omega_{1})/(1+\omega_{2})$.

Consider now the integral in (\ref{B4}) with $r_{2}\leq r\leq\srx r_{2}$. 
Since the factor 
$\exp\(-(\omega_{k_{1}}-\omega_{k_{2}})^{2}\frac{z}{4a(zx+ax^{4})}\)$
is smaller than one, the $k_{1}$, $k_{2}$ summations are bounded
by a product of ${\cal I}$ functions (\ref{cali}).
Hence, for $r\geq r_{2}$
\bea
0<F_{c}(x)&\leq& \exp\(-\frac{a^3r_{2}^{2}}{2}\)\int_{0}^{\infty}dv 
\int_{0}^{\srx r_{2}}dr\,\exp\(-\frac{a^3r^{2}}{2}\)\nonumber\\
&&\quad \times e^{a\omega_{1}r_{1}}
{\cal I}\(r_{1}+\frac{r}{\srx}+\frac{z}{2ax^{2}}\)
e^{a\omega_{1}r_{2}}{\cal I}\(r_{2}-\frac{r}{\srx}+\frac{z}{2ax^{2}}\)
\label{B7}
\eea
For ${\cal I}\(r_{1}+\frac{r}{\srx}+\frac{z}{2ax^{2}}\)$ we use the bound
${\cal I}\(r_{1}+\frac{r}{\srx}+\frac{z}{2ax^{2}}\)$$\leq  C\exp\(-a\omega_1
\(r_{1}+\frac{r}{\srx}+\frac{z}{2ax^{2}}\)\)$$\leq C\exp(-a\omega_1 r_1)$
since the argument becomes large as $x\to\infty$, whereas for
${\cal I}\(r_{2}-\frac{r}{\srx}+\frac{z}{2ax^{2}}\)$ we use the bound (see
the discussion leading to Eq.(\ref{as1}))
${\cal I}\(r_{2}-\frac{r}{\srx}+\frac{z}{2ax^{2}}\)$$\leq C
\(r_{2}-\frac{r}{\srx}+\frac{z}{2ax^{2}}\)^{-3/2}$
since the argument can become small when $r$ approaches the upper 
integration limit $\srx r_{2}$.
Thus
\bea
0<F_{c}(x)&\leq& C^2\exp\(-\frac{a^3r_{2}^{2}}{2}+a\omega_{1}r_{2}\)
\int_{0}^{\srx r_{2}}dr\,\exp\(-\frac{a^3r^{2}}{2}\)
\int_{0}^{\infty}{dz\over 
\(r_{2}-\frac{r}{\srx}+\frac{z}{2ax^{2}}\)^{3/2}}\non
&=& C^2 4ax^{5/2}\exp\(-\frac{a^3r_{2}^{2}}{2}+a\omega_{1}r_{2}\)
\int_{0}^{r_2} dr' {\exp\(-{a^3\over 2}x(r'-r_2)^2\)\over \sqrt{r'}}.
\label{B8}
\eea
The second line has been obtained by performing the $z$-integral 
and changing the integration variable $r$ to $r'=r_2-r/\srx$. 
This last integral in (\ref{B8}) is finite uniformly with respect to $x$ 
so that with $r_{2}=x/2-\nu_2$ the bound (\ref{B8}) tends to zero 
in a Gaussian way as $x\to\infty$. 
These last arguments can be reproduced to show that 
the integral with  $-\srx r_{1}\leq r\leq -r_{1}$ in Eq.(\ref{B4}) tends to
zero.

\end{document}